\documentclass{osa-article}

\journal{osac}

\usepackage{xcolor}
\usepackage[normalem]{ulem} 
\usepackage{floatrow}
\newfloatcommand{capbtabbox}{table}[][\FBwidth]

\articletype{Research Article}

\usepackage{algpseudocode}
\usepackage{soul}
\usepackage{amsmath,amsfonts}
\usepackage{graphicx}
\usepackage{textcomp}
\usepackage{xcolor}
\usepackage[english]{babel}
\usepackage{tikz}

\def\lB{\lambda_B}
\def\lS{\lambda_S}
\def\Tof{\emph{ToF}}
\def\Tacq{T_{acq}}
\def\bP{\mathbb P}

\begin{document}

\title{Histogram-less LiDAR through SPAD response linearization}

\author{Alessandro Tontini\authormark{1,2}, Sonia Mazzucchi\authormark{3}, Roberto Passerone\authormark{2},
Nicol\`o Broseghini\authormark{4} and Leonardo Gasparini\authormark{1}}

\address{\authormark{1}Fondazione Bruno Kessler, Trento, Italy\\
\authormark{2}Department of Information Engineering and Computer Science, University of Trento, Trento, Italy\\
\authormark{3}Department of Mathematics, University of Trento, Trento, Italy\\
\authormark{4}Department of Physics, University of Trento, Trento, Italy}

\email{\authormark{*}tontini@fbk.eu, sonia.mazzucchi@unitn.it,
roberto.passerone@unitn.it, nicolo.broseghini@studenti.unitn.it, gasparini@fbk.eu} 

\begin{abstract*}
  We present a new method to acquire the 3D information from a
  SPAD-based direct-Time-of-Flight (d-ToF) imaging system which does not
  require the construction of a histogram of timestamps and can withstand high flux operation regime.
  The proposed acquisition scheme emulates the behavior of a SPAD detector
  with no distortion due to dead time, and extracts the \Tof\ information by a
  simple average operation on the photon timestamps ensuring ease of
  integration in a dedicated sensor and scalability to large arrays.
  The method is validated through a comprehensive mathematical analysis, whose
  predictions are in agreement with a numerical Monte Carlo model of the
  problem.
  Finally, we show the validity of the predictions in a real d-ToF measurement
  setup under challenging background conditions well beyond the typical
  pile-up limit of 5\% detection rate up to a distance of 3.8~m.
\end{abstract*}

\section{Introduction and related work}
Spatial perception enabled by 3D imaging techniques is constantly gaining
interest for industrial~\cite{Sesta2022}, automotive~\cite{Hsiao2022}, space~\cite{Villa2021} and consumer applications.
As an example, the required level of self-awareness of autonomous driving vehicles demands for 3D imaging systems with high resolution and
high frame rate.
Unfortunately, these requirements are in conflict, constraining engineers to
performance-limiting tradeoffs.
In this paper we focus on Light Detection And Ranging (LiDAR)~/~direct-Time-of-Flight (d-ToF) measurement based on Single-Photon Avalanche
Diodes (SPAD), which is one of the most promising among active techniques~\cite{Morimoto2020}.
In a SPAD-based d-ToF measurement, the distance is extracted by measuring the
traveling time of a pulse of light projected from the source and reflected
back by the target to a detector that consists of a SPAD operating as a
photon-to-edge converter, coupled to a photon timestamping circuit (usually a
Time-to-Digital Converter (TDC) or a Time-to-Amplitude Converter
(TAC))~\cite{Li2021,TontiniGaspariniPancheriPasserone18TNS,Richardson2009}.
Due to hardware limitations, such as the detector dead-time and the
statistical nature of photons, together with the presence of uncorrelated
background light, a number of observations are usually accumulated into a
histogram memory to enhance the signal to noise ratio and extract the target
distance by means of signal processing techniques~\cite{sesta2023,Gyongy2022}.

The increased interest in Advanced Driver Assistance Systems (ADAS), where 3D vision is a pillar, is focusing the attention of researchers in developing techniques to increase the robustness of such systems against the effect of background light and against possible interference from similar devices. Concerning the problem of background light, one of the most effective and widespread technique is known as
\emph{photon coincidence}, which exploits the temporal proximity of photons
belonging to the reflected laser pulse to filter out unwanted background
photons which are more likely to be temporally
sparse~\cite{Niclass2013,Perenzoni2017}. With a different method, based on
a smart accumulation technique by Yoshioka et al.~\cite{Yoshioka2018}, the
signal to noise ratio (SNR) is increased by merging the information from
pixels observing similar regions of the scene. A different approach has been
recently proposed by Manuzzato et al.~\cite{Manuzzato2022}, where a per-pixel
circuit is able to automatically decrease the SPAD sensitivity reducing the
probability of saturation in case of high background intensities, favoring the
detection of laser photons. Yet another technique is known as time-gating,
where by means of a search procedure several sub-ranges of the scene are
measured, increasing the SNR at the expense of an increased
acquisition time~\cite{Padmanabhan2021}. Regarding the problem of mutual interference, Ximenes et al.~\cite{Ximenes2018} propose a spread-spectrum based technique, where the laser emission time is randomized from device to device, spreading
any other interference below the level of the signal of interest.
Another solution is based on the emission of two laser pulses whose temporal
relationship is different from device to device and used to actively discard
unwanted interference~\cite{Seo2021,Tontini2023}. 

These techniques, however, cannot cope with very high fluxes
because of two fundamental problems.
First, the histogram of timestamps appears distorted due to the dead-time of
SPAD detectors and timestamping circuits, which translates into a non-linear
response of the system to the incident flux of photons over time.
It is a widely held belief that the upper photon flux limit that still results
in a negligibly distorted histogram is given by 5\% of detected photons per
laser cycle~\cite{Becker2005}.
Second, the amount of data generated by the sensor is too large to scale to a large number of pixels.
Rapp et al.~\cite{Rapp2021} overcome the former problem by linearizing the
histogram through a Markov chain model of the photon detection times.
By using this method, the system can cope with up to 5 photoelectrons per
illumination cycle.
Gyongy et al.~\cite{Gyongy2020} mitigate the second problem by upscaling a low
resolution depth image based on a high resolution intensity image. A similar approach is used by Ruget et al.~\cite{Ruget2021}, where the native resolution of a depth image from a SPAD camera is increased by means of a deep neural network. Still the bottleneck is represented by the necessity to build and transfer a
histogram of timestamps for each pixel in the image from which the \Tof\ is
extracted.

One strategy to reduce the bandwidth requirement on the amount of data which
is transferred from the chip to the controller (usually an FPGA or $\mu$C) is
to integrate the histogram, or part of it, directly on chip.
Several solutions are proposed in the literature to integrate histogramming
capability on-chip~\cite{Lindner2018, MattioliDellaRocca2020, Zhang2019,
Hutchings2019, Kim2021, Stoppa2021, Kumagai2021, Gyongy2021,
Park2022, Zhang2022}, but despite the advantage in bandwidth performance
compared to other solutions where the histogram is built
off-chip~\cite{Niclass2013, Perenzoni2017,Ximenes2018, Padmanabhan2021,
Manuzzato2022}, still many limitations are present.
In general, the on-chip realization of either a \emph{partial} or \emph{full}
histogram requires additional area, which can be obtained by either
reducing the fill factor or by using expensive 3D-stacked solutions.

With the so-called \emph{partial} approach~\cite{Zhang2019, Kim2021, Seo2021,
Stoppa2021, Gyongy2021,Zhang2022, Park2022}, a reduced histogram memory is
available on-chip, therefore requiring a search procedure to identify the
location of the ensemble of histogram bins containing the laser peak.
In the literature, two techniques have been described to implement a
\emph{partial} histogram behavior.
With the so-called \emph{zooming} technique
\cite{Zhang2019,Kim2021,Seo2021,Park2022}, at the beginning of the
measurement the reduced set of histogram bins is spread across the entire
distance range.
By counting the number of photons detected on each bin, the reduced set of
histogram bins is concentrated over several iterations on a shorter range,
thus achieving the desired resolution on the estimated target distance.
With the other technique, called \emph{sliding}, the subset of histogram bins
is already set to the desired resolution, thus covering only a small portion
of the range.
Again by means of several iterations, the subset of
histogram bins slides across the entire range, and the number of photons at
each iteration is used to estimate the target distance.
Despite the intrinsic differences between the two methods, for both of them,
as outlined by Taneski et al.~\cite{Taneski2022}, a laser power penalty occurs
as more laser shots are required to find the laser peak location.

A \emph{full} histogram approach is possible with resource sharing, e.g., by
reallocating the same histogram circuitry to several pixels, as described by
Kumagai et al.~\cite{Kumagai2021}.
However, resource sharing resulted in only $\approx 27\%$ of the chip area
dedicated to the SPAD array, requiring a very high clock frequency of 500~MHz
and the design of a 3D stacked solution with a complex scanning illumination
approach.

In this paper we propose for the first time a robust method supported by a rigorous mathematical model to extract the 3D information from a set of acquired timestamps \emph{without the need to build a histogram}, which can also sustain high photon fluxes, enabling the possibility to operate beyond the standard limit of 5\% detection rate \cite{Becker2005} for pile-up distortion. The method can be implemented using only two registers and one accumulator for each pixel.
With such a low amount of resources, the per-pixel memory requirement is
reduced by more than 3 orders of magnitude compared to standard d-ToF
architectures (off-chip histogram)~\cite{Ximenes2018,Padmanabhan2021}, and by
a factor of $\simeq 5$ compared to architectures with on-chip,
full histogramming capability~\cite{Erdogan2017}.
We reach the goal in two steps.
First, we propose and evaluate an algorithm to efficiently extract the target
distance from a set of timestamps based on a simple on-the-fly average
operation, which does not require the allocation of a histogram memory.
Then, since the proposed algorithm works on the assumption that the detector response is linear, we present two acquisition schemes
that can be easily implemented on chip and emulate the behavior of a single
photon detector with no dead-time, providing the desired linear response to the input
flux of photons.
The proposed method is supported by analytical and numerical (Monte Carlo)
models and has been validated experimentally up to a distance of 3.8~meters
(mainly limited by the sensor used for characterization~\cite{Perenzoni2020})
under a background light equivalent to 85~kilolux and beyond the standard 5\%
rule for pile-up distortion~\cite{Becker2005}.

The paper is organized as follows.
In Section~\ref{sec_prelim_valid}, a typical d-ToF acquisition system with the
most important parameters of concern is described and preliminary
considerations on a histogram-less acquisition approach are provided with a
first validation by means of a Monte Carlo model.
In Section~\ref{sec_mathematic}, we provide the analytical proof of the
proposed acquisition method, while in Section~\ref{sec_acquisition_schemes} we
explain in detail two acquisition schemes which are needed to emulate the
response of a linear detector, together with a comparison against state-of-the art sensors in terms of memory requirement, scalability and tolerance to high background flux.
In Section~\ref{sec_measurements}, we provide measurement results from an
existing SPAD-based d-ToF sensor, showing that the proposed acquisition and
extraction schemes are capable of successfully computing the ToF without the
need for a histogram of timestamps.
Finally, we discuss future perspectives to advance the results found in this
work in Section~\ref{sec_conclusion}.

\section{Preliminary validation}
\label{sec_prelim_valid}
In this section, we present the principle of operation of a SPAD-based d-ToF
system with preliminary considerations and Monte Carlo simulations on the
histogram-less approach which will be further developed in the paper.

\subsection{Typical d-ToF operation}
A typical d-ToF image acquisition requires a pulsed laser and a time-resolved,
single photon image sensor with photon timestamping capabilities.
It works by sending periodic laser pulses and then measuring the arrival time,
or \emph{timestamp}, of the first detected photons reflected by the target
following each pulse.
Due to space and bandwidth limitations, the number of photon timestamps
generated per laser pulse is typically limited to one.

\begin{figure}[]
\captionsetup{width=1\textwidth}
\centerline{\includegraphics[width=\columnwidth]{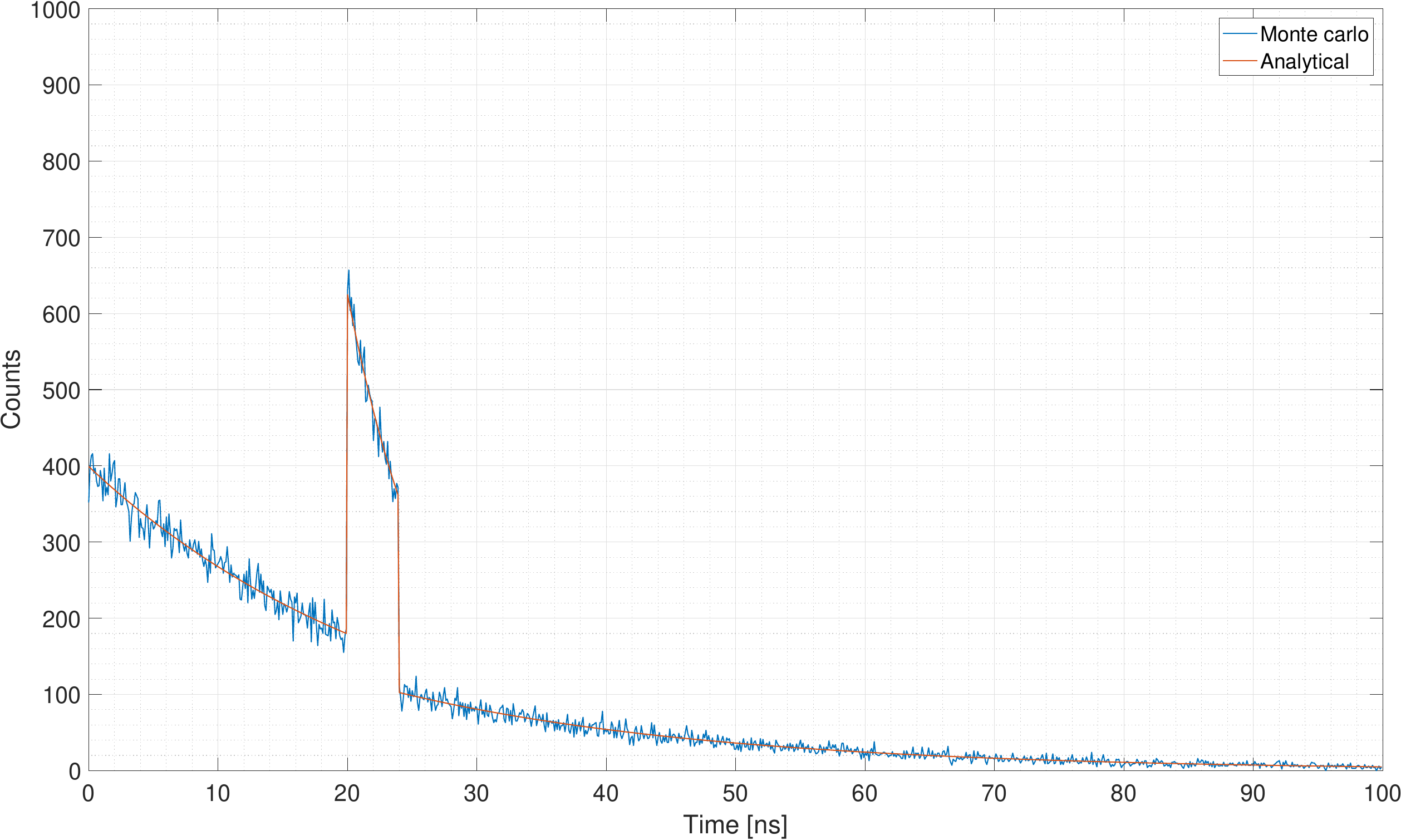}}
\caption{Simulated distribution of timestamps in a typical d-ToF system able
  to record 1 photon per acquisition, with a \Tof\ of 20~ns and a laser pulse
  duration $T_W$ of 4~ns. The histogram is composed of $10^5$ timestamps,
  with a bin size of 100~ps. Superimposed to the Monte Carlo simulation, we
  show also the analytical exponential distribution.}
\label{esempio}
\end{figure}

In principle, a single laser pulse, and thus a single timestamp, would be
sufficient to estimate the time of flight.
However, due to the presence of uncorrelated background events (from both
external light sources or internal SPAD Dark Count Rate (DCR)) and of shot
noise, the first detected photon may not be from the laser pulse, so that
several repetitions are needed to discriminate the different contributions.
To do so, the timestamps measured during the acquisition process are collected
in a histogram memory that records how many times each timestamp has been
observed.
This provides a convenient representation of the temporal distribution of the
arrival times, as shown in the example of Figure~\ref{esempio}.
In a system capable of acquiring only one photon (the first), the distribution
of the arrival times is a piece-wise exponential curve, where each segment is
described by:
\begin{equation}\label{eq0}
    P_i(t) = A_i \cdot e^{-\lambda_i \cdot t}
\end{equation}

whose \emph{intensity} (rate) $\lambda_i$ depends on the intensity of
background light, dark counts and laser echo.
Table~\ref{table_dToFparam} summarizes the most important parameters of the
detection process.

\begin{table}
\caption{List of parameters for a typical d-ToF acquisition.}
\centering
\begin{tabular}{c c c c }
\hline
\textbf{Parameter} & \textbf{Unit} & \textbf{Description}\\
\hline
$\lambda_B$ & $s^{-1}$ & Background events rate\\
$\lambda_S$ & $s^{-1}$ & Reflected laser events rate\\
$T_W$ & $s$ & Laser pulse duration\\
\Tof & $s$ & Time of flight\\
\hline
\end{tabular}
\label{table_dToFparam}
\end{table}

The \Tof\ is typically estimated from the histogram by finding the location of
its peak or of a sharp rising edge, which likely belongs to the reflected
laser pulse.
The histogram of timestamps contains all the relevant information to properly
estimate the time of flight, and represents the gold standard processing
technique in the field of SPAD-based d-ToF systems.
Unfortunately, the histogram requires a considerable amount of resources in
terms of memory, bandwidth and power, as it requires the readout of every
timestamp from the sensor for processing by an external controller (FPGA or
$\mu$C).
Even with the latest implementations where the histogram is available on-chip,
the required amount of resources is considerable.
As an example, a 10-m-range $128 \times 128$ LiDAR system with a 100~ps time
resolution and 8~bit histogram depth requires approximately 10~MB of memory.

\subsection{Histogram-less approach}
\label{sec_prelim_valid_MC}
Intuitively, if no background events are present, and neglecting the width of
the laser pulse, we could estimate the time-of-flight without the need to
build a histogram.
This can be achieved by simply calculating the average of the continuous
stream of laser-only timestamps.
To extend the above method to scenarios where background events are also
present, we need to eliminate their contribution to the average.
Again, intuitively, this can be accomplished by dividing the measurement into
two acquisitions.
The first is performed with the laser turned off, and is used to estimate the
contribution of the background light only, by computing the average $\bar
t_{bg}$ of the recorded timestamps.
The same operation is repeated in the second acquisition with the joint
contribution of background and laser timestamps, resulting in a total average
time $\bar t_{tot}$.
In principle, the time of flight should be proportional to the difference
between the two averages, with the contribution of the background canceling
out:
\begin{equation}\label{eq1}
  \Tof \propto \bar t_{tot} - \bar t_{bg}.
\end{equation}
This approach, however, relies on the superposition property which does not
hold, as SPADs are non-linear detectors.

\begin{figure}[]
\captionsetup{width=1\textwidth}
\centerline{\includegraphics[width=\columnwidth]{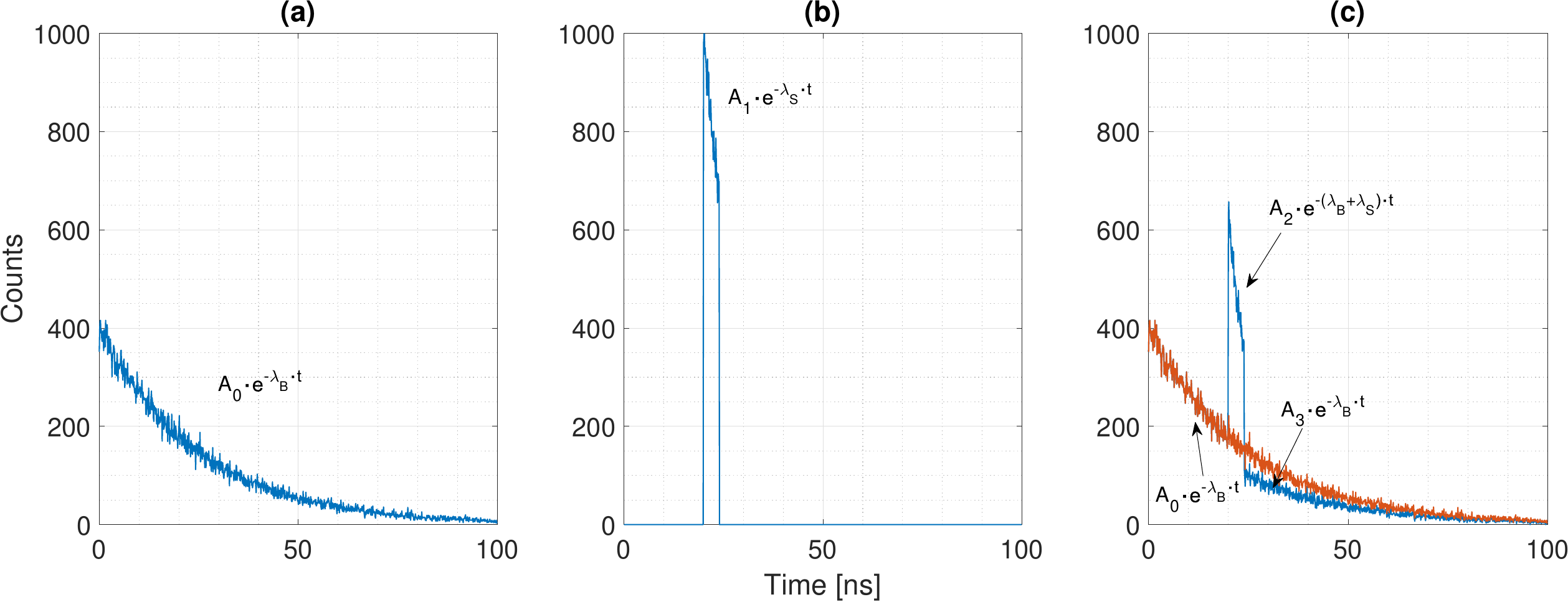}}
\caption{Simulated distribution of timestamps in a typical d-ToF acquisition.
  In (a) and (b), the distributions of background-only events (with rate
  $\lambda_B$) and laser-only events (with rate $\lambda_S$), respectively,
  are shown. In (c), the distribution of the combination of background and
  laser events is reported, graphically showing that the superposition
  property does not hold due to the non-linear behavior of the detection
  process. In particular, the portion of background events after the laser
  peak are underestimated, as only one photon per acquisition can be detected.
  For each contribution, the amplitude terms of the exponential ($A_0$, $A_1$,
  $A_2$ and $A_3$) are reported, with $A_3<A_0$ due to the SPAD
  non-linearity.}
\label{esempioScomposto}
\end{figure}

More specifically, the problem lies in the dead time of the detection process,
which depends upon the SPAD dead-time itself, on the limited bandwidth of the
timestamping circuit and also on the limited memory available, blinding the measurement channel for some time after each detection. With this limitation, the detection of a photon belonging to the laser echo does prevent a later
photon from being potentially detected, resulting in a distortion of the
statistics.
In particular, the amount of background photons which contribute to $\bar
t_{tot}$ in the second acquisition (with the laser turned on) is
underestimated.
This behavior can be observed on the histograms of
Figure~\ref{esempioScomposto}, where the distribution of timestamps in
different scenarios are compared, emphasizing the estimation error.
Furthermore, the greater the laser echo intensity, the higher the number of
background photons that are underestimated.

This can also be seen analytically by computing explicitly the cumulative
distribution function $F$ of the random variable $T$ associated to the first
photon detection time, defined as $F(t):=\bP(T\in [0,t])$.
If $t\in [0, \Tof]$ then the incoming photon necessarily belongs to the
background, hence 
$$ F(t)=\bP\left(T\in [0,t]\right)=1-e^{-\lB t}\, $$
If $t\in (\Tof, \Tof+T_W]$, then
\begin{align*}
    F(t)&=\bP(T\in [0,\Tof])+\bP(T\in (\Tof,t])\\
     &=1-e^{-\lB \Tof}+e^{-\lB \Tof}\left(1-e^{-(\lS+\lB)(t-\Tof)}\right) \,
\end{align*}
Finally, for $t>\Tof+T_W$ we have:
\begin{align*}
    F(t)&=\bP(T\in [0,\Tof])+\bP(T\in (\Tof,\Tof+T_W])+\bP(T\in (\Tof+T_W, t])\\
     &=1-e^{-\lB \Tof}+e^{-\lB \Tof}\left(1-e^{-(\lS+\lB)T_W}\right) +e^{-\lB
     \Tof}e^{-(\lS+\lB)T_W}\left(1-e^{-\lB(t-\Tof-T_W)}\right)\,
\end{align*}
The probability density $f$ of the distribution of $T$, where 
$F(t)=\int _{-\infty}^tf(u)du$, can be easily computed by means of the formula $f(t)=F'(t)$, yielding:
$$ f(t)=\begin{cases}
    \lB e^{-\lB t}  & t\in [0, \Tof]\\
    (\lS+\lB)e^{\lS \Tof}e^{-(\lS+\lB)t} & t\in (\Tof, \Tof+T_W]\\
    \lB e^{-\lS T_W}e^{-\lB t} & t >\Tof+T_W
\end{cases}$$
while the average first arrival time is given by
\begin{equation}
    {\mathbb E}[T]=\int_0^{+\infty}tf(t)dt=\frac{1-e^{-\lB \Tof}}{\lB}+ \frac{e^{-\lB \Tof}(1-e^{-(\lS+\lB) T_w})}{\lS+\lB}+\frac{e^{-\lS T_W}e^{-\lB (\Tof+T_W)} }{\lambda_B}\label{Eq-averagetime1}
\end{equation}
As shown in Equation~\eqref{Eq-averagetime1}, the average detection time
depends non-linearly on two parameters ($\lambda_S$ and \Tof).
This confirms that it is not possible to uniquely extract the \Tof\ from the
aforementioned acquisition method, since the average of the timestamps
acquired in the second acquisition, $\bar t_{tot}$, which is governed by the
probability of detection of laser photons, also depends on the $\lambda_S$
parameter, which depends not only on the \Tof, but also on the target reflectivity.
One could try to compensate for the error, however this requires measuring
also the intensity of the received laser light (which affects the error),
introducing an extra variable which is hard to estimate, invalidating the
procedure.

\begin{figure}[h]
\captionsetup{width=1\textwidth}
\centerline{\includegraphics[width=\columnwidth]{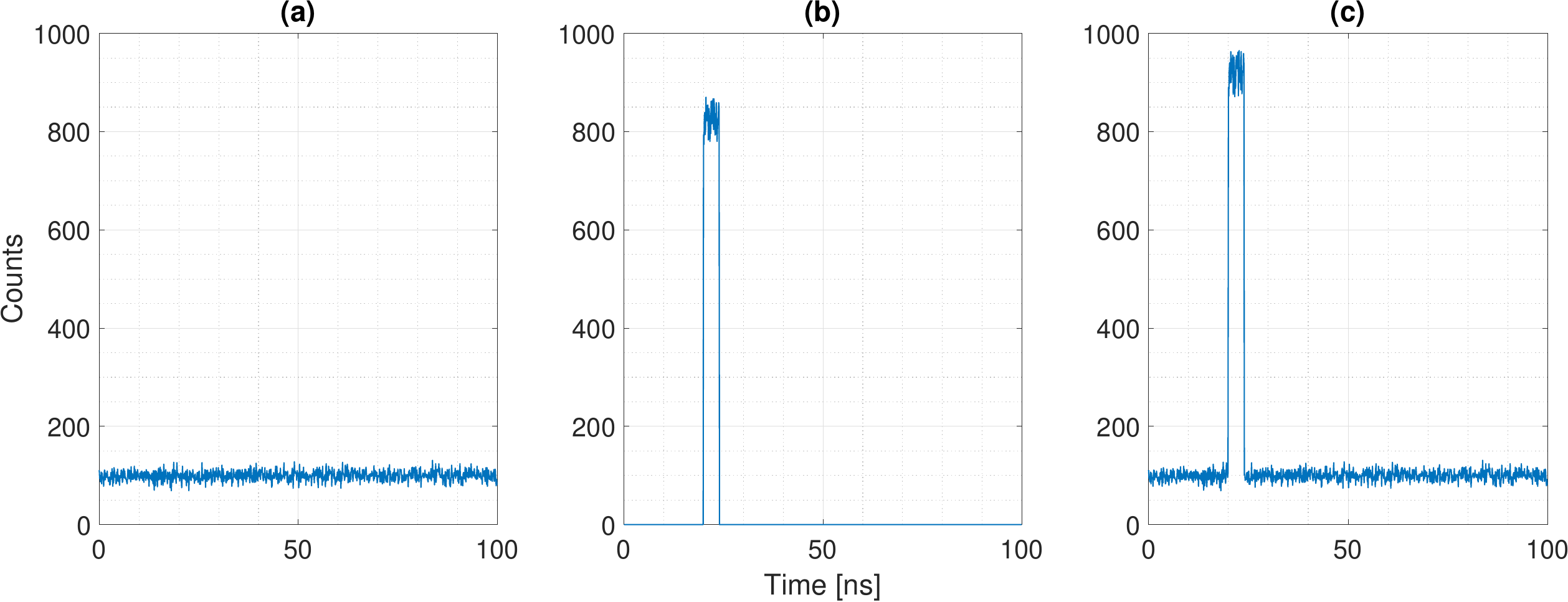}}
\caption{Distribution of timestamps obtained with a linear detection process, i.e., with no dead-time limitation. The distributions are uniform, since we are now considering the absolute arrival time of detected photons with respect to the beginning of the acquisition window. With this approach, it is possible to safely subtract the background contribution in (a) from the combined measurement in (c). As the superposition property holds in this case, there is no longer an under-weighting condition of background counts after the laser pulse peak in the histogram (c).}
\label{linearSPAD}
\end{figure}

Conversely, a linear detector, with no dead-time, is able to timestamp every
photon which falls within the acquisition window.
In this case, the histograms shown in Figure~\ref{esempioScomposto} become
linear in time, as shown in Figure~\ref{linearSPAD}.

One can then extract the \Tof\ with the proposed two-step procedure. In the first step, we measure the total number of \emph{background} events,
$N_{bg}$, and their average absolute arrival time, $\bar t_{bg}$.
In the second step, with the combination of both background and laser events,
we measure the \emph{total} number of events and their average absolute
arrival time, denoted $N_{tot}$ and $\bar t_{tot}$.
Because the superposition property holds, the difference $N_{tot} - N_{bg}$ is
equal to the amount of photons from the reflected laser source.
We can then extract the \Tof\ by properly weighting each average timestamp
measurement with the relative photon count contribution:

\begin{equation}\label{eq-tof-final-linearized}
N_{tot} \cdot \overline{t}_{tot} = N_{bg}\cdot\overline{t}_{bg}+(\Tof+\overline{t}_{l})\cdot(N_{tot}-N_{bg})\,
\end{equation}

where $\overline{t}_{l}$ is the average arrival time of the laser photons
referred to the laser emission time in the absence of background light (i.e.,
$\lB = 0$ and $\Tof = 0$).
The value $\overline{t}_{l}$ is a characteristic parameter of the laser
source, which can be experimentally estimated by means of an initial
calibration.
The proposed extraction method does not require the allocation of histogram
memory, and needs only two counters to store $N_{bg}$ and $N_{tot}$, and two
accumulators to compute $\bar t_{bg}$ and $\bar t_{tot}$ reducing the
memory requirements by more than three orders of magnitude compared to recent
long-range high-resolution d-ToF sensors~\cite{Ximenes2018,Padmanabhan2021}.
We can further reduce the amount of resources down to a single
accumulator, needed to store $\bar t_{tot}$, and two counters for $N_{bg}$ and
$N_{tot}$, because a constant background throughout the acquisition window
leads to an average background time $\bar t_{bg}$ of $T_{acq}/ 2$.
In this case, Equation~\eqref{eq-tof-final-linearized} turns into the simpler
form:

\begin{equation}\label{eq-tof-final-linearized-optimizd}
  N_{tot} \cdot \overline{t}_{tot} = N_{bg}\cdot\frac{T_{acq}}{2}+(ToF+\overline{t}_{l})\cdot(N_{tot}-N_{bg})\,
\end{equation}
which yields:
\begin{equation}\label{formula-Tof}
    \Tof = \frac{N_{tot}\bar t_{tot}-N_{bg}\frac{\Tacq}{2}}{N_{tot}-N_{bg}}-\overline{t}_{l}
\end{equation}

\begin{figure}[h]
\captionsetup{width=1\textwidth}
\centerline{\includegraphics[width=\columnwidth]{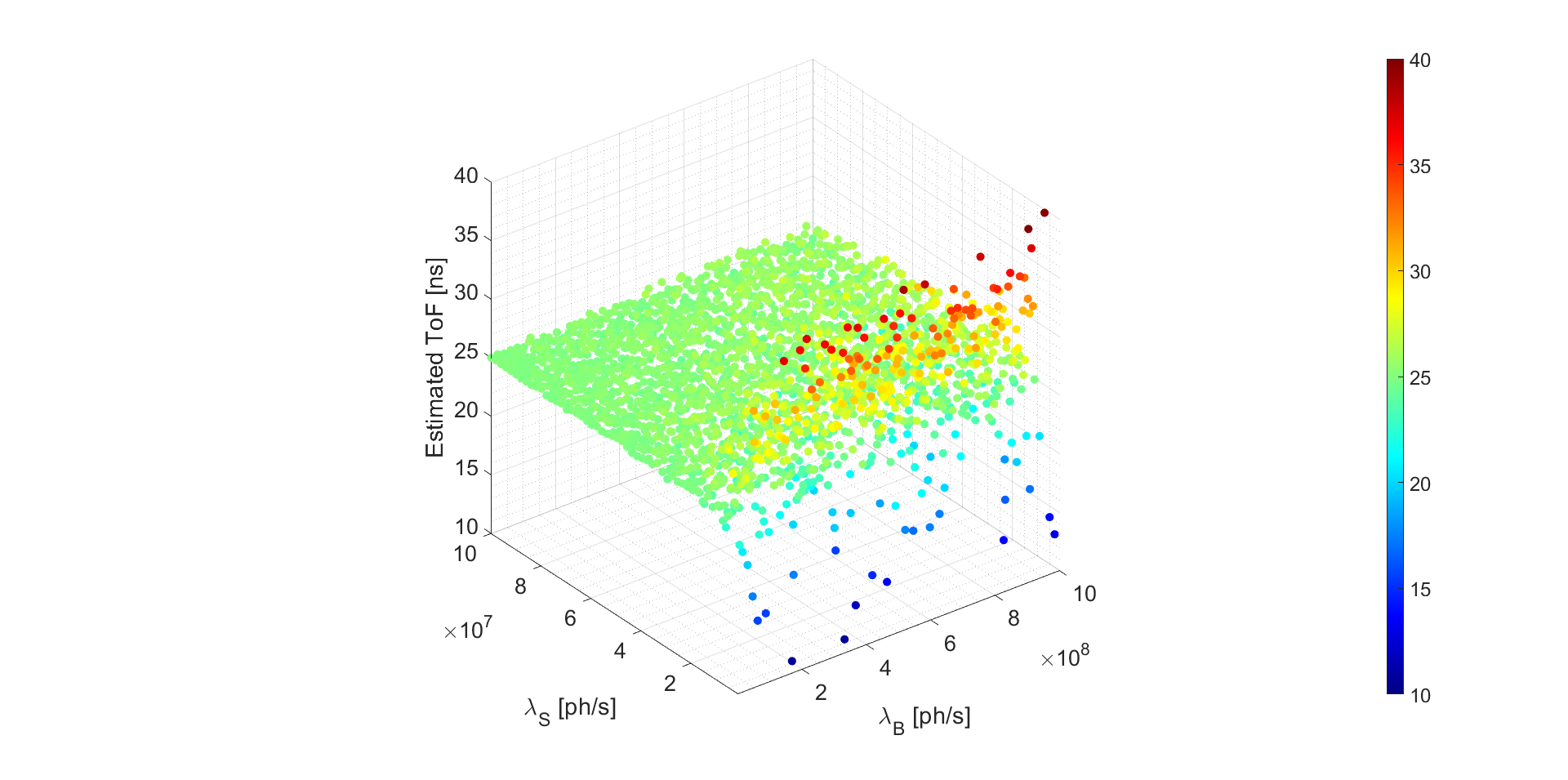}}
\caption{Preliminary Monte Carlo simulation results showing the \Tof\ computed
  with the proposed acquisition method with the hypothesis of an ideal linear
  detector. The \Tof\ can be properly estimated over a wide range of $\lambda_B, \lambda_S$ pairs.}
\label{TOF_linearSPAD}
\end{figure}

We have simulated this extraction method with a Monte Carlo
simulator~\cite{Tontini2020} by sweeping the parameters $\lambda_S$ and
$\lambda_B$ in the range $[10^{6}-10^{8}]$ and $[10^{5}-10^{9}]$, respectively.
For each pair of $\lambda_S$ and $\lambda_B$ values, $10^4$ measurements have
been acquired with the \Tof\ value set to 25~ns.
The resulting \Tof, obtained from Equation~\eqref{eq-tof-final-linearized}, is
shown in Figure~\ref{TOF_linearSPAD} with the correct estimation over a wide
range of $\lambda_B, \lambda_S$ pairs, failing only when the
$\lambda_S/\lambda_B$ ratio is too low even for a classic histogram-based
approach.

In the next section, we provide a rigorous mathematical analysis which
proves the results briefly introduced with Equation~\eqref{formula-Tof} from
the underlying statistical distribution of photons.

\section{Mathematical analysis}
\label{sec_mathematic}
This section proves the validity of the method described above analytically.
In the following, we shall denote the duration of the acquisition window by $T_{acq}$,
assume the laser echo to entirely occur within it, i.e., $T_{acq} \gg \Tof +
T_W$, and denote the time-dependent intensity of the laser pulse by the
function $\lS:[0, T_W] \to {\mathbb R}$.
The flux of photons can be modeled by a counting process $(N_t)_{t \in [0,
T_{acq}]}$ obtained as the sum of two independent Poisson processes:
$(N^B_t)_{t \in [0, T_{acq}]}$, with intensity $\lambda_B$, describing the
background flux of photons, and $(N^S_t)_{t \in [\Tof, \Tof + T_W]}$,
describing the signal.
In particular, the process $(N^S_t)_{t\in [\Tof, \Tof+T_W]}$ is modeled
through an inhomogeneous Poisson process with time dependent intensity
$(\tilde\lambda_S(t))_{t\in [\Tof, \Tof+T_W]}$ given by $\tilde\lambda
_S(t):=\lS(t-\Tof)$.
Hence, the overall flux of photons reaching the SPAD is given by an
inhomogeneous Poisson process $(N_t))_{t\in [0, \Tacq]}$ with varying
intensity $\lambda$ given by
\[
  \lambda (t) = \begin{cases}
    \lB              &  t\in [0, \Tof)\cup [\Tof+T_W, \Tacq] \\
    \lB+\tilde\lS(t) & t\in [\Tof, \Tof +T_W)
  \end{cases}
\]
Hence, we can prove that, given $n$ photons detected in the interval $I = [0,
\Tacq]$, the $n$ detection times are independent and distributed on $[0,
T_{acq}]$ with a distribution density function $f : [0, T_{acq}] \to {\mathbb
R}$ given by

\begin{equation}\label{densityTn-gen}
  f(t) = \begin{cases}
    \frac{\displaystyle \lB}{\displaystyle \lB \Tacq + \int_{0}^{T_W}\lS(u)du} & t \in [0, \Tof) \cup [\Tof + T_W, \Tacq] \\
    & \\
    \frac{\displaystyle \lB + \lS(t-\Tof)}{\displaystyle \lB \Tacq + \int_{0}^{T_W}\lS(u)du} & t \in [\Tof, \Tof + T_W)
\end{cases}
\end{equation}

Indeed, by considering a partition $\{I_j\}_{j=1,\dots, m}$
of the interval $I = [0, \Tacq]$, the independence of the increments of the
inhomogeneous Poisson processes $(N_t)_{t \in [0, \Tacq]}$ yields
\begin{equation}\label{eq-1-conditional-cumulative-gen}
  {\mathbb P}(N(I_1)=n_1, \dots, N(I_m)=n_m \mathbin{|} N(I)=n) =
  \frac{\displaystyle \prod_{j=1}^m e^{-\lambda_{I_j}}\frac{(\lambda_{I_j})^{n_j}}{n_j!}}
       {\displaystyle e^{-\lambda_{[0,\Tacq]}}\frac{\left(\lambda_{[0,\Tacq]}\right)^{n}}{n!}}
\end{equation}
where, for an interval $I = [a,b] \subset {\mathbb R}$ we adopted the notation
$\lambda _I := \int_a^b \lambda(t) dt$ and we assumed $\sum_{j=1}^m n_j = n$.
By a straightforward computation, the right hand side
of~\eqref{eq-1-conditional-cumulative-gen} can be written as
\begin{equation*}
 { \mathbb P}(N(I_1)=n_1, \dots, N(I_m)=n_m \mathbin{|} N(I)=n) =
  \frac{n!}{n_1!\cdots n_m!}\prod_{j=1}^m \left(\frac{\lambda_{I_j}}{\lambda_{[0, \Tacq]}}\right)^{n_j}\, 
\end{equation*}
the latter being equivalently obtained in terms of $n$ independent and
identically distributed continuous random variables $T_1, \ldots, T_n$ with
density $f$ given by~\eqref{densityTn-gen}.
In other words, the  arrival times $\{T_i\}_{i=1, \ldots, n}$ of the $n$
photons provide a statistical sample for the distribution
\eqref{densityTn-gen}.\\
The distribution~\eqref{densityTn-gen} has a mean $\mu$ given by
\begin{equation}\label{mean-gen}
    \mu=\int_0^{\Tacq}tf(t)dt =\frac{1}{\lB \Tacq+\int_{0}^{T_W}\lS(t)dt}\left( \frac{\lB\Tacq^2}{2}+\Tof \int_{0}^{T_W}\lS(t)dt+ \int_{0}^{T_W}t\lS(t)dt\right)
\end{equation}
Clearly, since $\mu$ is a linear function of $\Tof$, it can easily be inverted. By denoting with $\alpha$ the ratio
\begin{equation}\label{par-alpha}
    \alpha:= \frac{\lB \Tacq}{\displaystyle \lB\Tacq + \int_0^{T_W}\lS (t)dt}\, 
\end{equation}
we get an equation providing the time-of-flight as a function of the other characteristic parameters of the process:
\begin{equation}\label{formula-tof-gen}
\Tof=\frac{1}{1-\alpha}\left(\mu-\alpha\frac{\Tacq}{2}\right)-\overline{t}_{l}\,
\end{equation}
where $\overline{t}_{l} = \frac{\int_0^{T_W} t \lS(t)dt}{\int_0^{T_W} 
\lS(t)dt}$ is the average arrival time of the laser photons
referred to the laser emission time in the absence of background light (i.e.,
$\lB = 0$ and $\Tof = 0$).

If $n$ photons are detected within the interval $[0, \Tacq]$, the sample mean
$\bar t_{tot}$ of their detection times provides an unbiased estimator for
$\mu$.
The main issue with this approach is the estimation of the parameter $\alpha$,
which depends on $\lB$ and $\lS$, the latter being affected by a high level of
uncertainty since it is related to the intensity of the laser echo.
Because the total number of photons detected in the time interval $[0, \Tacq]$
is a Poisson random variable with average $\lB\Tacq + \int_0^{T_W}\lS(t)dt$
and, analogously, the number of background photons detected in the time
interval $[0, \Tacq] $ is a Poisson random variable with average $\lB\Tacq$,
we can estimate both parameters by observing a realization of both processes.
More precisely, let us first switch off the laser source and collect the
number $N_{bg}$ of photons arriving during the interval $[0, \Tacq] $, then
let us switch on the laser source and collect the total number $N_{tot}$ of
photons arriving during the interval $[0, \Tacq]$.
The observed values $N_{bg}$ and $N_{tot}$ are respectively an estimate for
$\lB \Tacq$ and $\lB\Tacq+\int_0^{T_W}\lS(t)dt$, while their ratio is an
estimate $\hat \alpha:= \frac{N_{bg}}{N_{tot}}$ for the parameter $\alpha$
defined in \eqref{par-alpha}.
By replacing the parameters $\mu$ and $\alpha$ with their estimates $\bar t_{tot}$ and $\hat \alpha$ in formula~\eqref{formula-tof-gen}, we obtain the
following estimator for $\Tof$:
\begin{align}\label{estimate1}
    \widehat{\Tof} & = \frac{1}{1-\hat\alpha}\left(\bar t_{tot}-\hat\alpha\frac{\Tacq}{2}\right)-\overline{t}_{l}\\
    & = \frac{\displaystyle N_{tot}\bar t_{tot}-N_{bg}\frac{\Tacq}{2}}{N_{tot}-N_{bg}}-\overline{t}_{l}\,
\end{align}
which coincides with \eqref{formula-Tof}.

\section{Acquisition schemes}\label{sec_acquisition_schemes}
The simulation results obtained in Section~\ref{sec_prelim_valid_MC} are based
on the assumption that the photon detection process is ideal, i.e., with no
dead-time and with a linear response over the incoming flux of photons.
In a real-world scenario, however, detectors are limited by the dead-time
between subsequent detections, resulting in a non-linear response.
To implement the proposed extraction method, we propose a novel SPAD
acquisition scheme which emulates the behavior of a linear detector.
More in detail, we propose two ways to obtain a linearized SPAD response from
a real SPAD.
Both methods are based on the assumption that the underlying statistical
processes are stationary and ergodic.
In particular, we assume that there are no major fluctuations of the
characteristic parameters of the process during the acquisition time.
Similarly to an equivalent-time sampling oscilloscope, both methods rely on
repeating the observation multiple times to emulate the response of a SPAD
detector with no dead time.
In Section~\ref{subsec:acq1} and~\ref{subsec:acq2}, we describe the
working principle of each method and propose a possible implementation.
Then, in Section~\ref{subsec:acq_description}, we provide the mathematical
proof that both acquisition methods are capable of correctly sampling the
distribution of photon arrival times.

\subsection{Acquisition scheme \#1: Acquire or discard}
\label{subsec:acq1}
\begin{figure}[]
\captionsetup{width=1\textwidth}
\centerline{\includegraphics[width=\columnwidth]{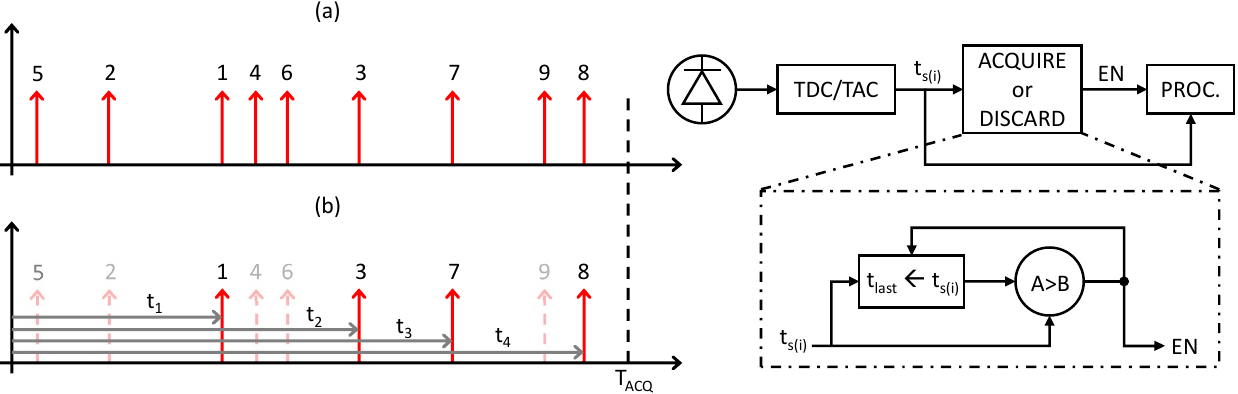}}
\caption{Example of SPAD response linearization with the \emph{acquire or
  discard} acquisition method.
  In (a), each photon arrival time is represented by a red arrow and the order
  of arrival is indicated.
  The first run starts with the acquisition of photon~\#1, resulting in
  timestamp $t_1$.
  Photon~\#2 is discarded, since its arrival time is earlier than photon~\#1.
  The next recorded information comes instead from photon~\#3, which is later
  than photon~\#1, and sets the new minimum time.
  The run proceeds with the same criteria resulting in the stream of
  photon arrival times $t_1$, $t_2$, $t_3$ and $t_4$ from photons~\#1,
  \#3, \#7 and \#8, which is a single realization of the emulated response of
  the linearized SPAD detector.
  On the right, a principle schematic is proposed, showing the lightweight
  usage of resources, with only one comparator and one register required on
  top of the processing circuit.
  The \emph{acquire or discard} acquisition method is simple but inefficient,
  as most of the photons arrival times are discarded, resulting in longer
  acquisition times.}
\label{AcquireOrDiscard}
\end{figure}

The first acquisition scheme relies on a simple (albeit inefficient) mechanism
which requires no additional resources in terms of the SPAD driving circuit.
The acquisition works over multiple runs, each requiring multiple observations.
The first timestamp of every run is considered valid, memorized, and used to
increment either $N_{bg}$ or $N_{tot}$, depending on the current phase of the
acquisition, and update $\overline{t}_{tot}$.
Then, in the next observations, timestamps are considered valid, and used to update
the algorithm parameters, only if they are higher than the largest previous
timestamp, otherwise they are discarded.
This procedure is repeated until the end of the acquisition window $T_{acq}$
is reached (no photon detected), thus covering a complete acquisition, and
concluding the run.
The process is then repeated multiple times to increase statistics.
Figure~\ref{AcquireOrDiscard} shows an example of a run, including all the
discarded events, and a possible implementation.
While the implementation is straightforward, the method is inefficient because
the majority of the detected photons may end up being discarded.

\subsection{Acquisition scheme \#2: Time-gated}
\label{subsec:acq2}
\begin{figure}[] \captionsetup{width=1\textwidth}
  \centerline{\includegraphics[width=\columnwidth]{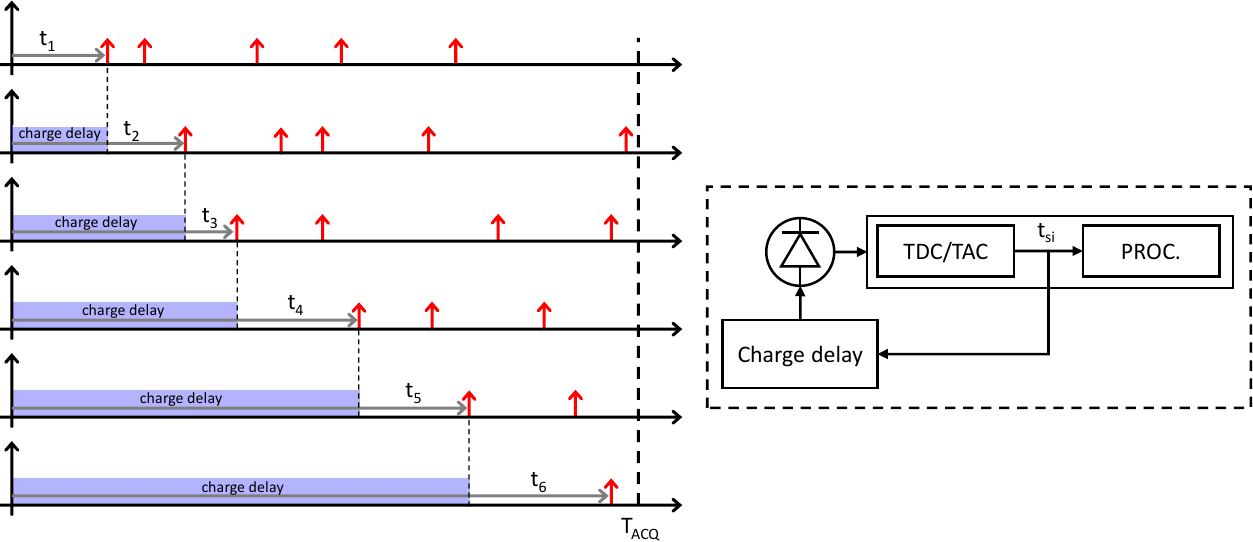}}
\caption{Example of SPAD response linearization with the \emph{time-gated}
  acquisition method. With this approach, no photon timestamp is discarded
  thanks to the delayed activation of the SPAD for each timing measurement.
  During the \emph{charge delay} phase, the SPAD front-end is forced OFF, thus
  photons can not be detected. For each measurement, the first timestamp is
  detected and used to increment either $N_{bg}$ or $N_{tot}$ and update the
  average time $\overline{t}_{tot}$. At the same time, the \emph{charge delay}
  phase value is updated accordingly for the next measurement. As opposed to
  the \emph{acquire or discard} method, more hardware resources are needed to
  build the delay element which controls the activation of the SPAD.}
\label{fig:time_gated}
\end{figure}

The time-gated acquisition scheme works by delaying the activation of the SPAD
to start from the previously acquired timestamp, until the end of the
acquisition window is reached. With this approach, there is no need to discard
timestamps, allowing for a faster acquisition.
This, however, comes at the expense of a more complex hardware implementation,
which needs a time-activated gating scheme, for instance using a programmable
delay line.
An example of acquisitions is shown in Figure~\ref{fig:time_gated}, together
with a possible implementation.

\subsection{Mathematical description}\label{subsec:acq_description}
\begin{figure}[]
\captionsetup{width=1\textwidth}
\centerline{\includegraphics[width=\columnwidth]{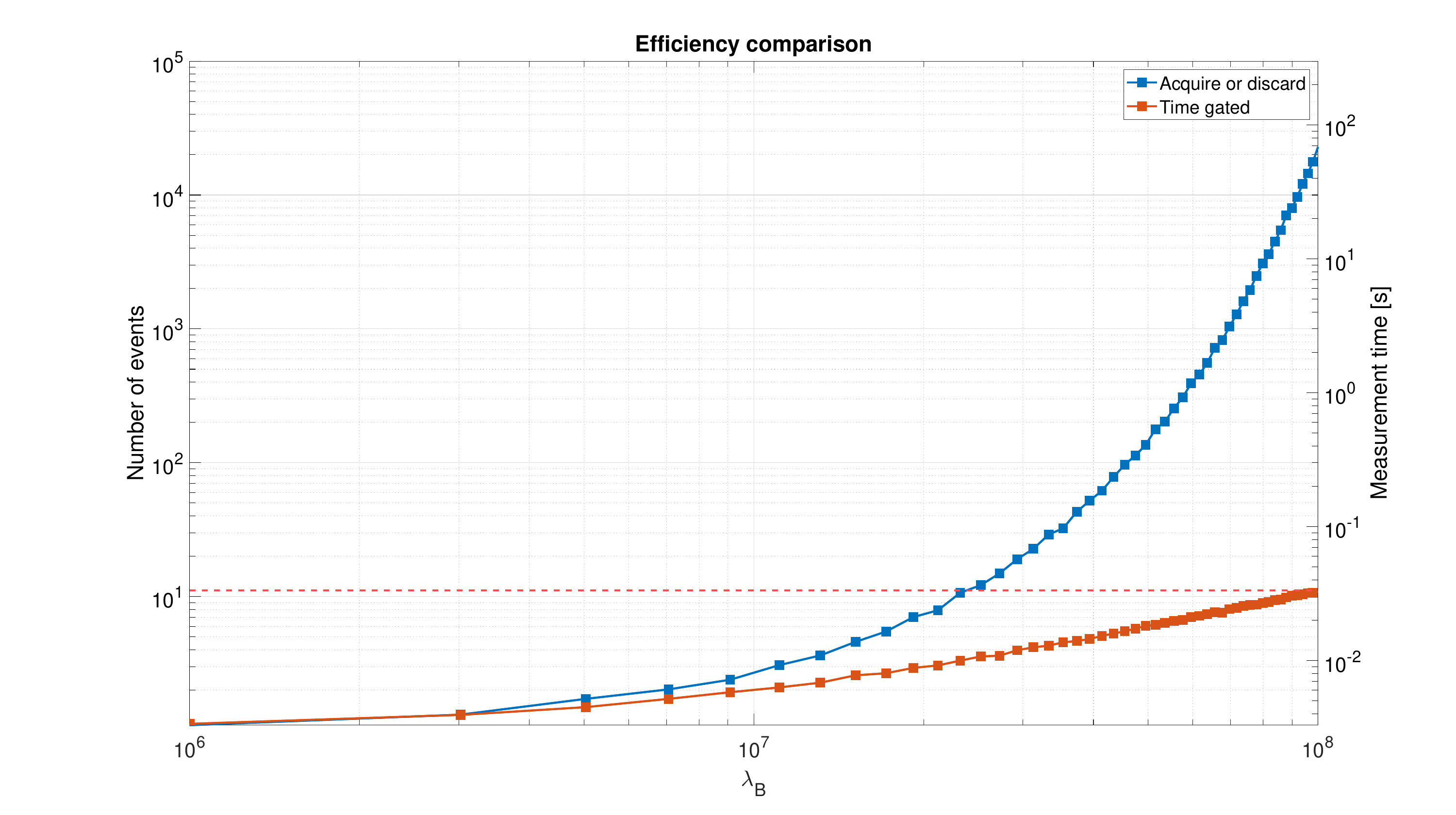}}
\caption{Result of Monte Carlo simulation comparing the two proposed acquisition schemes in terms of efficiency for increasing values of background light flux ($\lambda_B$) in the range $[10^6,10^8]$~ph/s. On the left axis, we show the number of timestamps required to linearize the SPAD response over an acquisition window $T_{acq}$ of 100~ns. On the right axis, we show the total time required for the two methods to collect $N=3\cdot10^4$ measurements, in order to average the linearized response of the SPAD N times. The horizontal line indicates a limit of $\approx33.3$~ms, for an equivalent operation frame-rate of 30~FPS. With the \emph{time gated} scheme, the required frame rate can be guaranteed over the entire range of background light flux, while considering the \emph{acquire or discard} scheme, the maximum sustainable flux is limited to $\approx2.4\cdot10^7$~ph/s.}
\label{speedComparison}
\end{figure}

While providing the same result, it is clear that the implementation cost and
the performance of the two acquisition schemes are different.
With the \emph{acquire or discard} scheme, almost no hardware modification is
required to an already existing SPAD sensor.
However, because of the decimation process, the efficiency of the acquisition
could be very low.
This also depends on the intensity of the incoming flux of photons: the higher
the intensity, the higher the probability to have smaller timestamps which
block the detection process.
On the other hand, the \emph{time-gated} scheme requires a delay line and the
SPAD-gating, but the efficiency is much higher since no decimation process
occurs.
To show the difference in terms of efficiency of the two proposed acquisition
schemes, we run a Monte Carlo simulation with background light flux in
the range $[10^6,10^8]$~ph/s with the aim to linearize the SPAD response over an
acquisition window $T_{acq}$ of 100~ns.
As shown in Figure~\ref{speedComparison}, at the highest flux of
$10^8$~photons/sec, the amount of timestamps to be acquired to cover the
acquisition window $T_{acq}$ for the \emph{acquire or discard} scheme is more
than 3 orders of magnitude higher compared to the more efficient \emph{time
gated} scheme.
From the simulation, we can also identify the maximum number of measurements
which can be executed by the two acquisition schemes to sustain an operation
frame rate of 30~Frames Per Second (FPS).
The \emph{time-gated} scheme can average the linearized SPAD response up to $N = 3
\cdot 10^4$ times over the whole range of background light flux.
On the other hand, with the same number $N$ of acquisitions, the \emph{acquire
or discard} scheme can only support up to $\approx 2.4 \cdot 10^7$~ph/s of
background flux.

From a mathematical point of view, both acquisition methods allow sampling
the correct distribution of the photon arrival times $(T_n)_{n\geq 1}$. Let $T_n$ denote the occurrence time of the $n-$th event, i.e., the arrival of
the $n$-th photon.
This can be defined as the infimum of the set of times $t$ such that the
number of arrivals $N_t$ in the interval $[0,t]$ is greater than or equal to
$n$:
\[
  T_n :=\inf\{t \geq 0\, :\, N_t\geq n\}, \qquad n\geq 1
\]
By definition, $N_{T_n} = n$.
We can therefore derive an equivalent representation of the random variables
$T_n$.
Indeed, for $n\geq 2$, the time $T_n$ of occurrence of the $n-$th event can be
obtained as the infimum of the set of times $t$ greater than $T_{n-1}$ (the
time of occurrence of the $(n-1)$-th event) such that the increment
$N_t-N_{T_{n-1}}$, the number of events occurring in the interval
$(T_{n-1},t]$, is greater than 1.
Therefore

\begin{align*}
T_1 &:=\inf\{t \geq 0\, :\, N_t\geq 1\}\\
 T_n &:=\inf\{t \geq T_{n-1}\, :\, N_t-N_{T_{n-1}}\geq 1\}, \quad n\geq 2
\end{align*}

which corresponds to the results of the acquisition schemes described
previously.

\subsection{Comparison with state-of-the-art}
\begin{figure}[]
\captionsetup{width=1\textwidth}
\centerline{\includegraphics[width=\columnwidth]{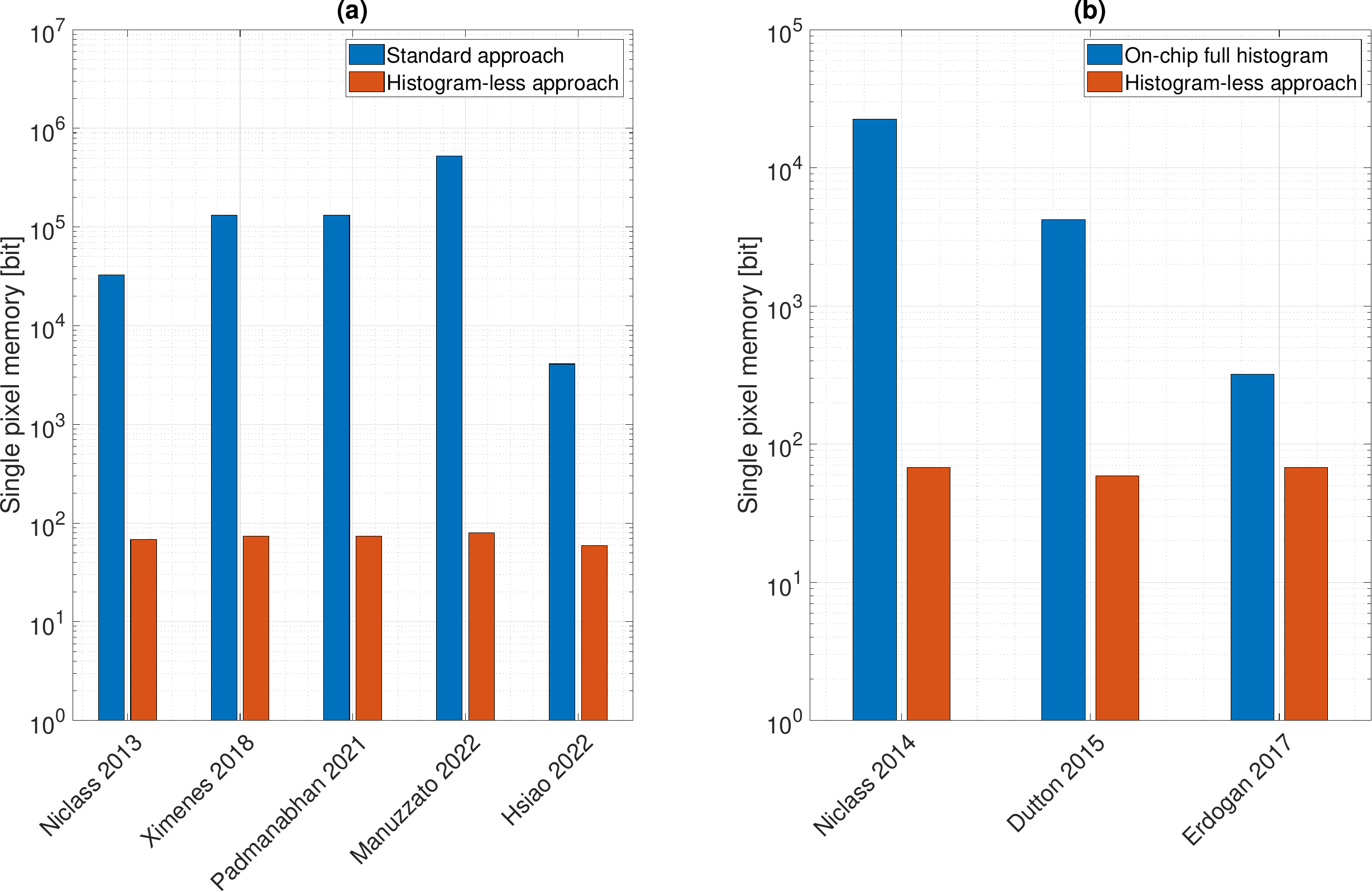}}
\caption{Comparison of the amount of per-pixel memory required by our
  histogram-less acquisition method against histogram-based d-ToF sensors. In
  (a), we consider standard sensors where every timestamp is read out and the
  histogram is built off-chip~\cite{Niclass2013, Ximenes2018,
  Padmanabhan2021,Manuzzato2022,Hsiao2022}.  In (b), we consider sensors with
  full on-chip histogram capability~\cite{Niclass2014,Dutton2015,Erdogan2017}.}
\label{memoryComparison}
\end{figure}

In this section, we compare our histogram-less acquisition method with
state-of-the-art SPAD-based LiDAR sensors in terms of memory requirement,
scalability and tolerance to high background light flux.
For all comparisons in this section, we consider for our method 16~bits of
counters depth (i.e., up to 65535 counts for $N_{bg}$ and $N_{tot}$,
respectively) and then 3 times the number of TDC bits (1xTDC bits required for
the TDC word itself, and then 2xTDC bits to properly size the accumulator
memory).
First, we compare against standard sensors, i.e., sensors that require the raw
timestamps to be read out to build the necessary histogram of timestamps
off-chip.
To provide a fair comparison, we do not consider the sensor resolution, which
changes from chip to chip, but only the amount of memory required to build the
histogram for one pixel.
In the works we consider for our comparison~\cite{Niclass2013, Ximenes2018,
Padmanabhan2021,Manuzzato2022,Hsiao2022}, we extrapolate the total amount of
per-pixel memory based on the number of reported TDC bits and on an 8-bit
histogram depth for all of them.
We then compare our solution to sensors that offer full on-chip histogram
capability~\cite{Niclass2014,Dutton2015,Erdogan2017}.
Also in this case, we consider the amount of memory reported in each work
necessary to build the histogram of timestamps for one pixel.
Results are reported in Figure~\ref{memoryComparison}, where an average memory
reduction factor of $\approx2129$ and $\approx136$ for standard and full
on-chip histogram sensors, respectively, is obtained.
More comparison details, including minimum and maximum memory reduction
factors are reported in Table~\ref{table_memoryComparison}.

Similarly to our method, partial histogram
approaches~\cite{Zhang2019,Hutchings2019,Kim2021,Zhang2022} are also quite
effective in reducing the memory requirements.
Nevertheless, our approach not only outperforms them with a memory reduction
that ranges from 67\%~\cite{Hutchings2019} to 3\%~\cite{Zhang2022}, but also performs better in many other important aspects.
In fact, unlike previous work, our approach does not have any of the following
needs:
\begin{itemize}
    \item The need to find the laser peak in time using a zooming or a sliding
      search procedure, which is at the basis of every partial histogram
      approach~\cite{Taneski2022}.
    \item The need to share hardware resources (TDCs, memory) among pixels in
      the same column~\cite{Zhang2019} or in the same
      cluster~\cite{Zhang2022}, to reduce the area usage.
    \item The need for area consuming processors to manage the algorithm
      underlying the partial histogram technique and that can only be
      implemented using advanced 3D integrated technologies with single-pixel
      access~\cite{Hutchings2019}.
\end{itemize}
All these translate into higher measurement time and laser power penalty as
more acquisitions are needed than a standard full-histogram approach~\cite{Taneski2022}, and
higher costs.

Our method, given the very limited amount of required memory resources, is
also advantageous in terms of scalability to higher sensor resolutions, and
also in terms of range extension.
As an example, a standard histogram-based sensor with 15-bit TDC requires
memory to store up to 32767 histogram bins per pixel.
If the measurement range is doubled, the additional TDC bit results in an
increase of 100\% on the memory requirement.
Conversely, with our approach, the amount of memory increase to double the
range is limited to only $\approx$~3.9\%.

Concerning the tolerance to high background light flux, both
detection processes can sustain very high flux regimes, with a limit
determined by the finite resolution $T_{TS}$ of the timestamping circuit.
This limit translates to the requirement $\frac{1}{\lambda_B + \lambda_S} \gg
T_{TS}$, i.e., having a low probability that more than one photon fall into
the same time bin.
By considering a threshold on this probability, we can extract
the maximum flux of photons $\lambda_{max}$ which can be sustained by our
detection process.
The probability to have more than one photon per time bin is expressed as
$P(n>1) = 1-e^{-\lambda_{max}\cdot T_{TS}}\cdot(1+\lambda_{max}\cdot T_{TS})$.
By setting a threshold of less than 1\%, and considering $T_{TS}=100$~ps, the
maximum photon flux that can be sustained is equal to $\lambda_{max}\simeq
1.48\cdot 10^9$~ph/s.
Compared to the maximum flux required by a standard system which must comply
with the 5\% rule, and with the hypothesis of an acquisition window $T_{acq}$
of 100~ns, our detection process can sustain a photon flux $\simeq$~3000 times
higher.

\begin{table}
\caption{Minimum, average and maximum memory reduction factor of the proposed histogram-less acquisition method against standard d-ToF sensors (off-chip histogram)~\cite{Niclass2013, Ximenes2018,
Padmanabhan2021,Manuzzato2022,Hsiao2022} and sensors with on-chip full histogram capability~\cite{Niclass2014,Dutton2015,Erdogan2017}.}
\centering
\begin{tabular}{c c c c c c c c}
\hline
\multicolumn{4}{c}{\textbf{Standard sensors}} & \multicolumn{4}{c}{\textbf{Full on-chip histogram}} \\
\hline
Min. & Avg. & Max. & ~ & ~ & Min. & Avg. & Max. \\
\hline
69.4~\cite{Hsiao2022} & 2129.5 & 6553.6~\cite{Manuzzato2022} & ~ & ~ & 4.7~\cite{Erdogan2017} & 135.9 & 331.3~\cite{Niclass2014}\\
\hline
\end{tabular}
\label{table_memoryComparison}
\end{table}

\section{Measurement results}\label{sec_measurements}
The proposed acquisition scheme has been validated with measurements using
real data from an existing single-point SPAD-based d-ToF sensor, with an
architecture similar to the one from Perenzoni et al.~\cite{Perenzoni2020},
which in addition offers on-chip histogramming capability. The sensor is fabricated in a standard 150~nm CMOS process with the SPAD technology developed in the work by Xu et al.~\cite{Xu2017}.
The histogram features 1024 bins with 10-b depth, and a TDC resolution of
100~ps.
The SPADs are enabled synchronously with the beginning of the acquisition
window and the first measured timestamp for each acquisition increments the
corresponding histogram bin.
After a user-selectable number of acquisitions, the histogram is read out and
unpacked.
Then, the unpacked data is shuffled to recover a realization vector of the
arrival times of the detected photons.

\begin{figure}[]
\captionsetup{width=1\textwidth}
\centerline{\includegraphics[width=\columnwidth]{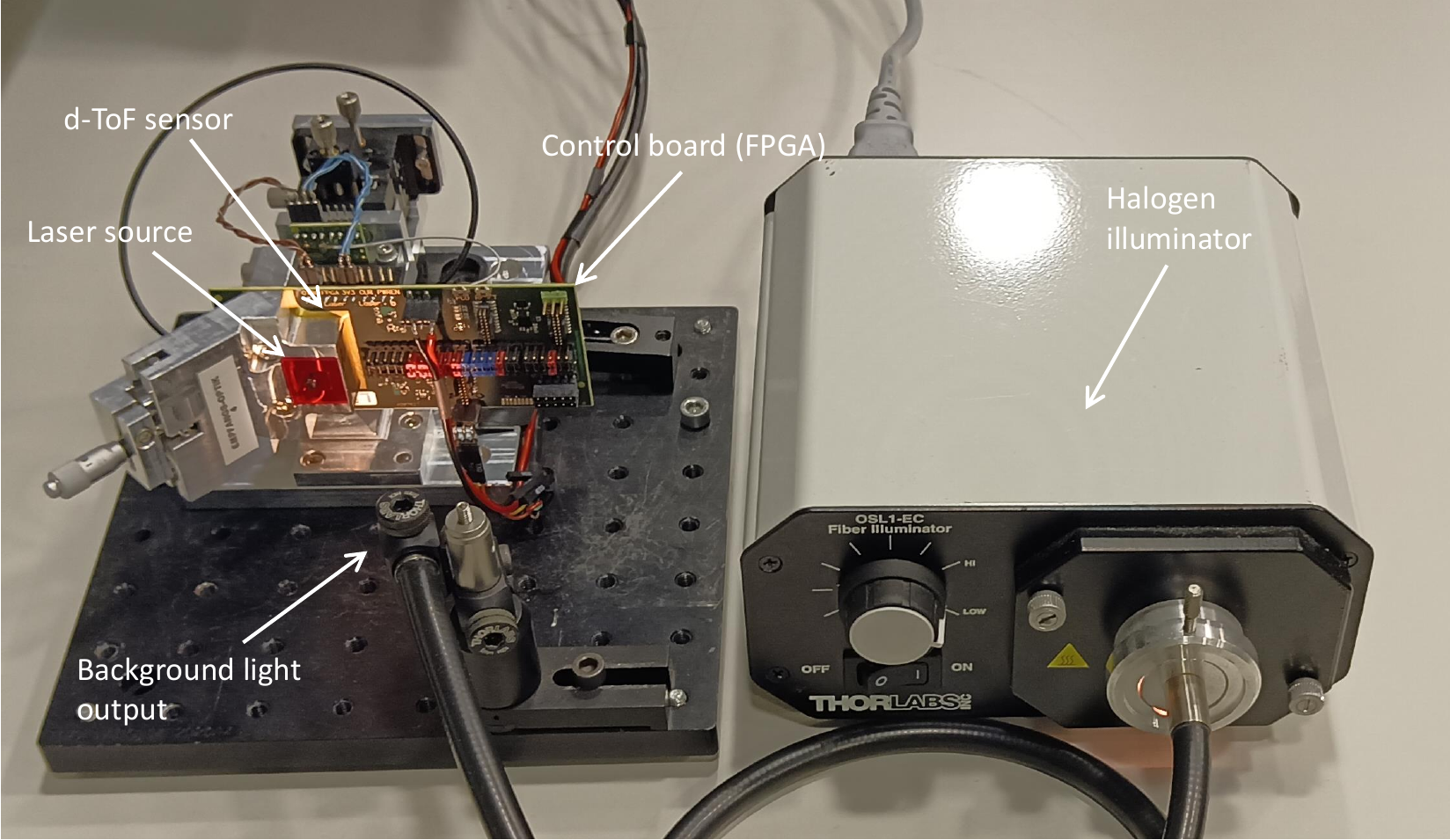}}
\caption{Measurement setup with the FPGA control board, d-ToF system and
  halogen illuminator for the generation of background light pointed directly
  toward the sensor.}
\label{setup}
\end{figure}

Background events were generated by means of a $\approx$180~W fiber-coupled
halogen illuminator pointed directly toward the sensor, while a black matte panel
with low $\approx$10\% reflectivity was selected as target, with a distance
range from 1~m up to 3.8~m.
A picture of the setup is shown in Figure~\ref{setup}, with indications on the
main components.

First, we focus on the validation of the linearization behavior of the
proposed acquisition scheme by considering only background light.
Then, we consider the combination of background and laser together, as in a
real scenario, and we compute the ToF with the proposed histogram-less
acquisition scheme.

\subsection{Preliminary considerations}
As we base our measurements on the re-engineering of an existing d-ToF sensor, preliminary considerations are needed before providing further details on measurement results. The sensor measures the arrival time of the first detected photon for each laser pulse, as described in Section~\ref{sec_prelim_valid}, which is stored in an on-chip histogram memory. Since the sensor measures the arrival time of the first photon, the statistical distribution is exponential, thus we are considering \emph{relative} arrival times. A statistically valid realization of the incoming timestamps is obtained by unpacking and randomly shuffling the content of the histogram memory. The obtained realization is a vector of \emph{relative} arrival times, which is the starting point of our measurement analysis.

In Section~\ref{sec_acquisition_schemes}, we described two possible
acquisition schemes.
The \emph{acquire or discard} scheme, even though is intrinsically
inefficient, can be straightforwardly used with our dataset as it requires no
hardware modification over the already existing SPAD-based d-ToF system.
The \emph{time-gated} scheme, while more efficient, requires a time-gating
circuit which is not implemented in our sensor.

The first set of measurements focuses on background events only.
In this case, there is a single source of events with intensity $\lambda_B$,
so we can apply the \emph{time-gated} scheme by computing the cumulative sums
of timestamps to obtain \emph{absolute} arrival times from \emph{relative}
ones.
On the other hand, when events from background and laser are combined, as in a
real measurement scenario, it is not possible to mimic the behavior of the
\emph{time-gated} scheme by means of the cumulative sum operation.
In that case we rely on the \emph{acquire or discard} scheme.

\subsection{Measurements with background light only}
\begin{figure}
\captionsetup{width=1\textwidth}
\centerline{\includegraphics[width=\columnwidth]{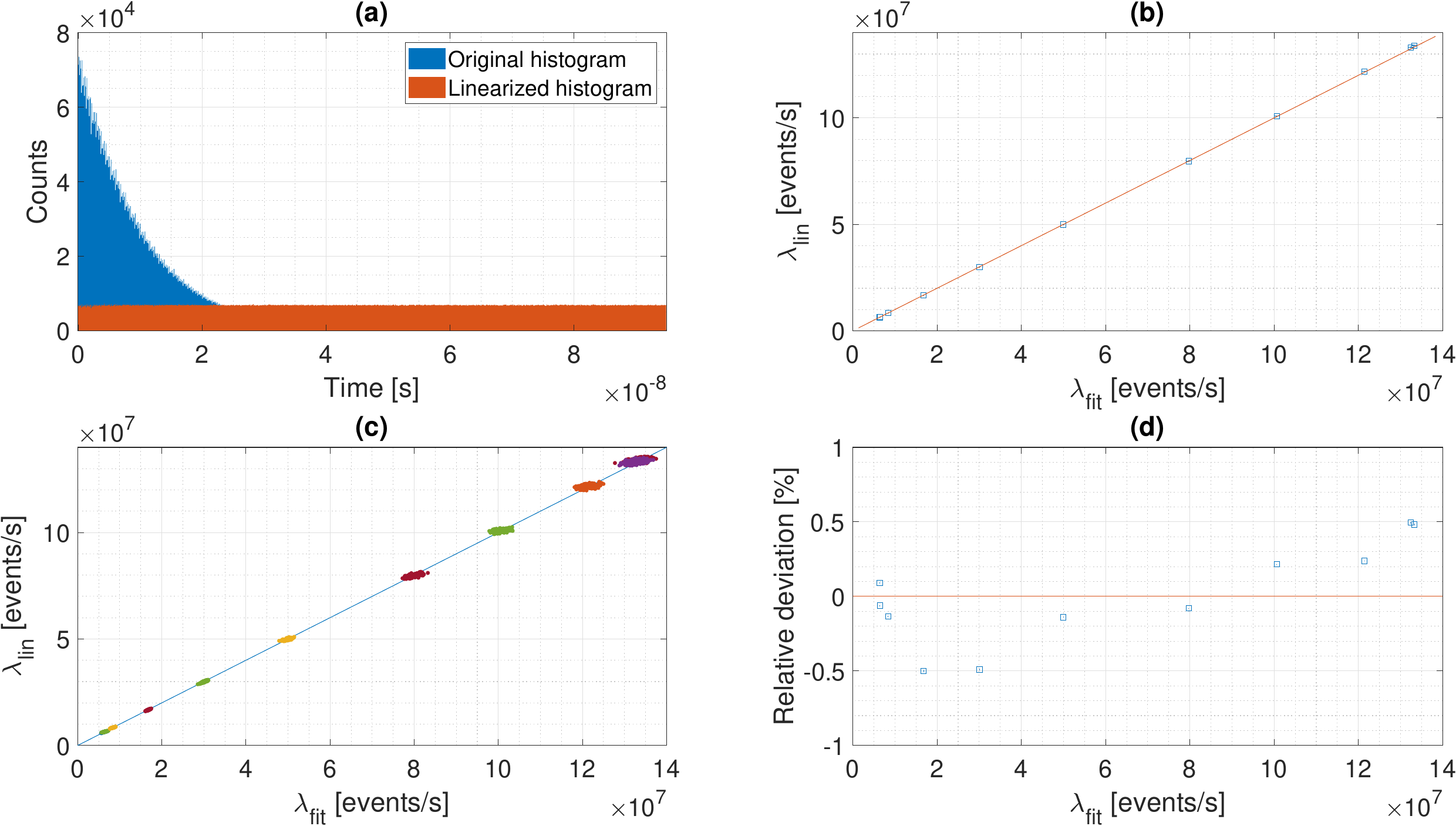}}
\caption{Linearization of the SPAD response with background events only. For each value of background flux, $8\cdot10^6$ timestamps are acquired from
  the sensor. In (a), an example of linearized histogram is
  shown together with the original one (exponentially distributed) for a background flux of $\approx100\cdot10^6$~events/s. In (b), we show the flux of background events estimated from the
  linearized histogram of timestamps, $\lambda_{\textit{lin}}$ against
  the flux estimated from an exponential fit on the original histogram of
  timestamps, $\lambda_{\textit{fit}}$. In (c), for each value of background
  flux, the entire dataset was split in 200 subsets to analyze the
  homogeneity of the linearization process, while in (d), the relative
  deviation from the background flux measured from the original
  histograms is shown, used as a reference, demonstrating a relative deviation below $\pm0.5$~\% over all
  data subsets.}
\label{BGmeas_1}
\end{figure}

We set the intensity of background light from a minimum of $\approx 6.5\cdot
10^6$ up to $\approx 133\cdot 10^6$ events/s.
This is the rate of events at the output of the SPAD, which therefore takes
into account all physical parameters of concern of a typical d-ToF
system~\cite{Tontini2020}.
Considering an acquisition window of 100~ns, specific from the
sensor~\cite{Perenzoni2020}, the equivalent average number of detections
within $T_{acq}$ equals $\approx 0.65$ and $\approx 13.3$ for the minimum and
maximum background light intensity, respectively.
In both cases, this is much higher than the conventional limit of 5\%
events~\cite{Becker2005} (13 and 266 times higher, respectively), showing the
high resistance of our method against pile-up distortion.
By considering the equation which links the intensity of background events,
$\lambda_B$, with the physical parameters of the system~\cite{Tontini2020}, it
is possible to derive the equivalent background illumination level, in
kilolux, up to a maximum of $\approx$~85~kilolux.
Measurement results are shown in Figure~\ref{BGmeas_1}, showing a relative
deviation from the reference background intensity extracted from the
exponential fit of the original histogram of less than $\pm0.5$~\% over the
whole range of values.

\subsection{Measurements with background and laser light and extraction of the \Tof}
\begin{figure}
\captionsetup{width=1\textwidth}
\centerline{\includegraphics[width=\columnwidth]{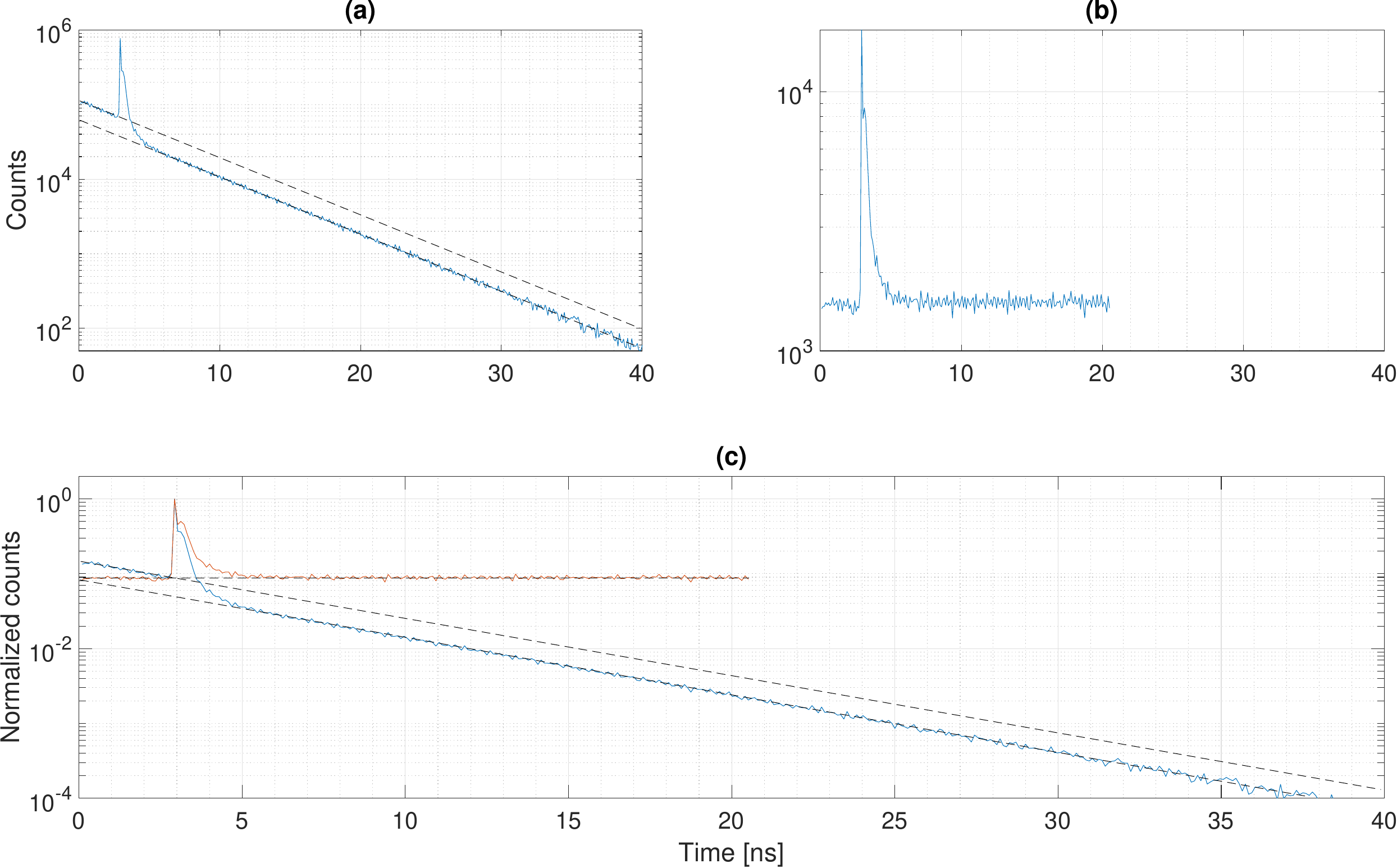}}
\caption{Qualitative measurement showing the linearization process of the
  proposed acquisition scheme. In (a), the original histogram of timestamps is
  shown in logarithmic scale, where the drop of counts which occurs after the
  laser peak is clearly visible. In (b), the histogram obtained with the
  \emph{acquire or discard} scheme proves the efficacy of the linearization
  process, which fully compensates for the non-linearity of the detector. The
  length of the linearized histogram is shorter than the original dataset, as
  we decided to stop the linearization earlier to reduce the data loss which
  naturally occurs with the \emph{acquire or discard} scheme. Due to the
  intrinsic inefficiency of this scheme, the histogram peak in (b) is
  attenuated by $\approx34$~dB with respect to the original dataset in
  (a). In (c), the two histograms are shown together after normalization.}
\label{BG_las_meas}
\end{figure}

Our first goal is to show that the underestimation of background counts which
occurs in a standard d-ToF system can be completely recovered with our
acquisition scheme.
This is demonstrated in the first measurement, displayed in
Figure~\ref{BG_las_meas}, which compares a traditional acquisition with the
\emph{acquire or discard} scheme, qualitatively showing the linearization
process by means of the linearized histogram of timestamps.

\begin{figure}
\captionsetup{width=1\textwidth}
\centerline{\includegraphics[width=\columnwidth]{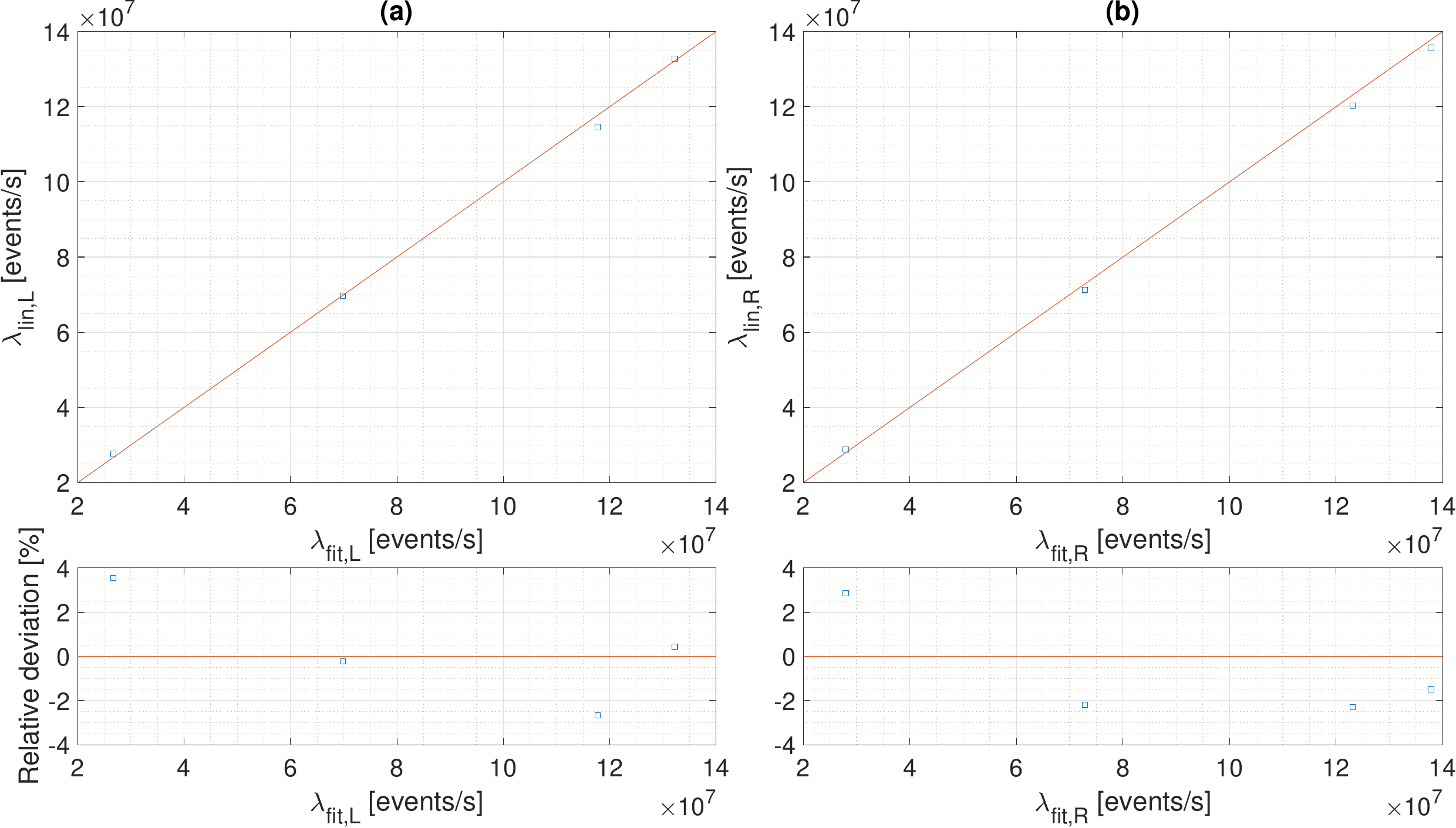}}
\caption{Quantitative characterization of the linearization process
  considering four different values of background flux, from
  $\approx27.6\cdot10^6$~events/s up to $\approx133\cdot10^6$~events/s with a target
  distance of 2.5~m. For each value of background flux, $2.5\cdot10^6$ timestamps
  are acquired from the sensor. In (a), the relationship between the
  background fluxes computed before the laser peak is shown, where
  $\lambda_{\textit{fit},L}$ comes from an exponential fit on the original
  histogram, while $\lambda_{\textit{lin},L}$ comes from the linearized
  histogram. In (b), the same relationship is shown but considering the
  portion of background events after the histogram peak. For each portion, the
  relative deviation of the flux extracted from the linearized histogram
  of timestamps is shown, demonstrating an estimation error below $\pm4$~\%
  over the whole range. The application of the
  \emph{acquire or discard} acquisition scheme results in a data reduction
  factor of $\approx7.5$ and $\approx165$ for the minimum and maximum
  background light flux, respectively.}
\label{BG_las_all}
\end{figure}

We then quantitatively evaluated the linearization process by estimating the
intensity of background light from both portions of the histogram, i.e.,
before and after the laser peak.
For this characterization, we used the $\approx$10\% reflectivity target (black matte panel) at
2.5~m distance from the sensor.
The results are depicted in Figure~\ref{BG_las_all}, showing a relative
deviation from the ground truth (estimated from an exponential fit on the
original dataset) below $\pm4$~\%.

\begin{figure}
\captionsetup{width=1\textwidth}
\centerline{\includegraphics[width=\columnwidth]{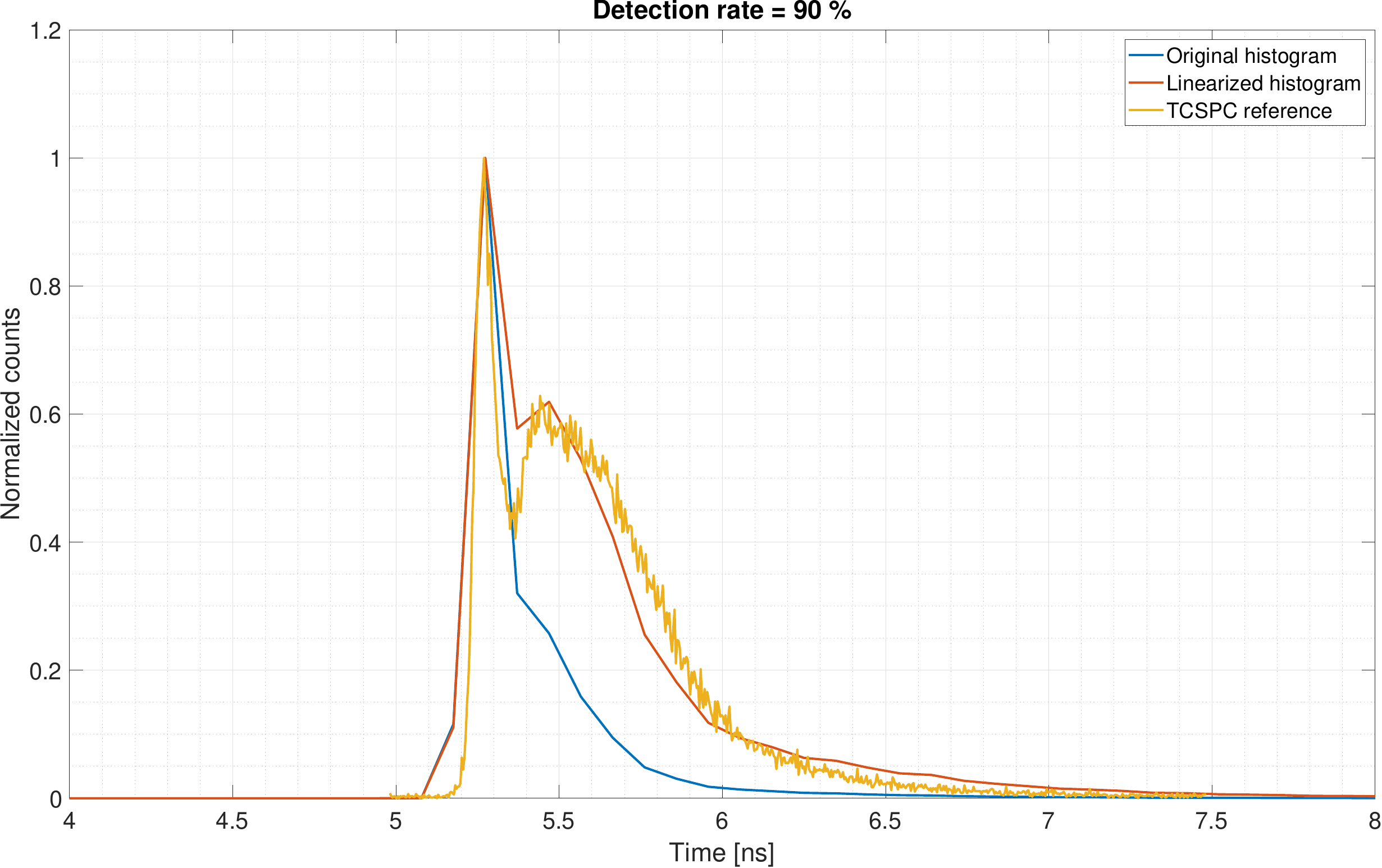}}
\caption{Characterization of the behavior of the proposed SPAD linearization method under strong pile-up conditions. 
The histogram obtained from the linearized vector of timestamps is compared against the original histogram (built from the detection of the first arrival time) and against a reference measurement obtained with a conventional TCSPC setup. In the histograms obtained from our sensor timestamps, the bin width is 100~ps, while the reference measurement from the TCSPC setup has 4~ps timing resolution. The proposed SPAD linearization method allows us to recover the full shape of the laser envelope even if the detection rate is 18 times higher than the conventional limit of 5\%.}
\label{pileUpMeas}
\end{figure}

In a different measurement, we verify the resistance of the proposed SPAD
linearization method against pile-up distortion.
To do so, we acquire several timestamps from the reflected laser pulse with a
detection rate of 90\%, which is 18 times higher than the conventional limit
of 5\%.
The results, shown in Figure~\ref{pileUpMeas}, proves the efficacy of our
linearization method in challenging pile-up conditions where a standard sensor
would fail.
A reference measurement acquired with a conventional Time-Correlated Single
Photon Counting (TCSPC) setup is shown as reference.

\begin{figure}
\captionsetup{width=1\textwidth}
\centerline{\includegraphics[width=\columnwidth]{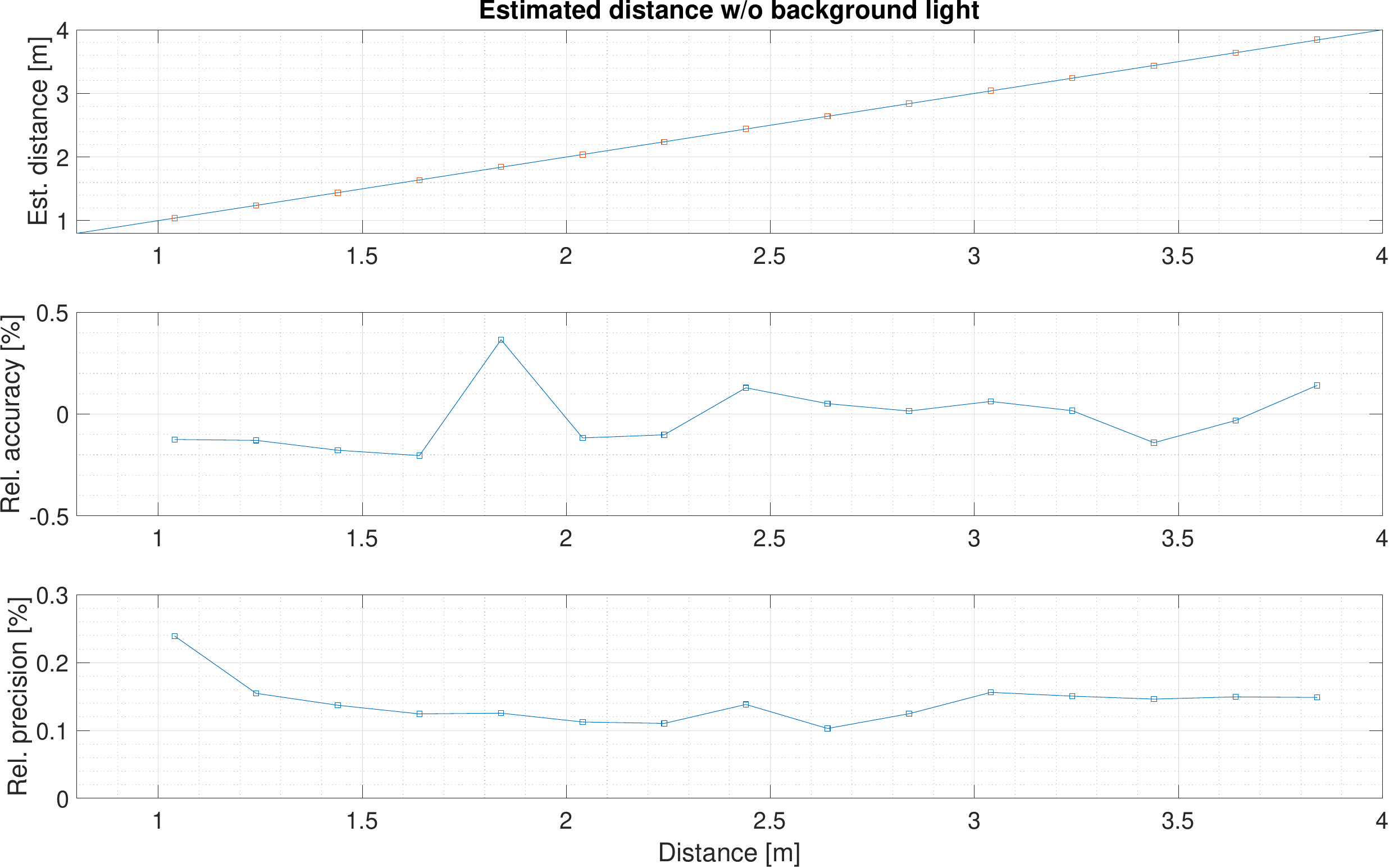}}
\caption{Measurement results with no background light, showing the \Tof\ extracted without the need to build a histogram of timestamps. The relative accuracy is below $\pm0.5\%$, while the relative precision is below $0.25\%$ for all measurements.}
\label{ToFnoBG}
\end{figure}

\begin{figure}
\captionsetup{width=1\textwidth}
\centerline{\includegraphics[width=\columnwidth]{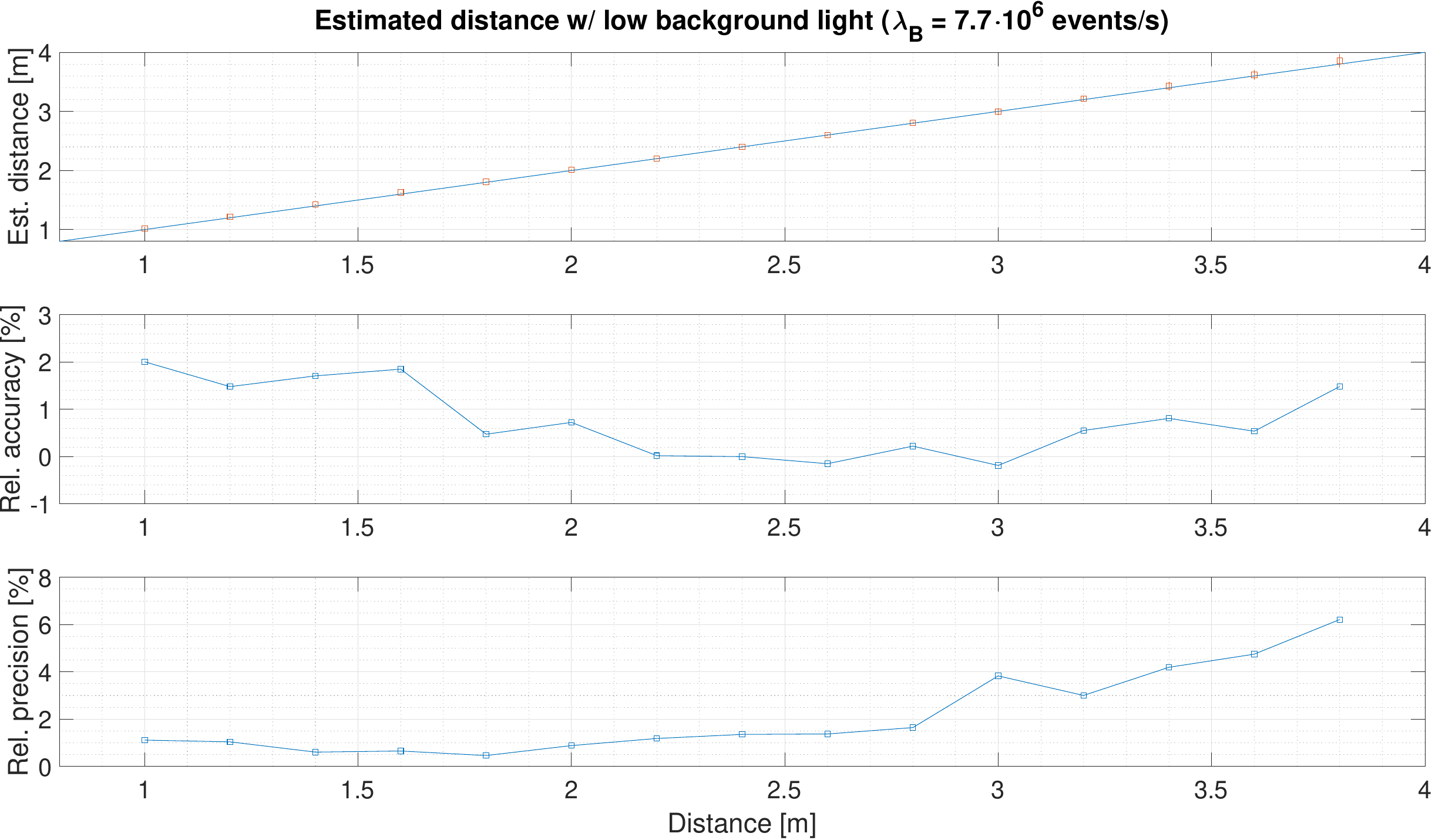}}
\caption{Measurement results with low background light flux ($\lambda_B=7.7\cdot10^6$~events/s), showing the
  extracted \Tof\ without the need to build a histogram of timestamps. The relative accuracy is in the range [-0.2,2]~\%, while the worst relative precision is $6\%$ at the highest distance of 3.8~m.}
\label{ToFlowBG}
\end{figure}

\begin{figure}[]
\captionsetup{width=1\textwidth}
\centerline{\includegraphics[width=\columnwidth]{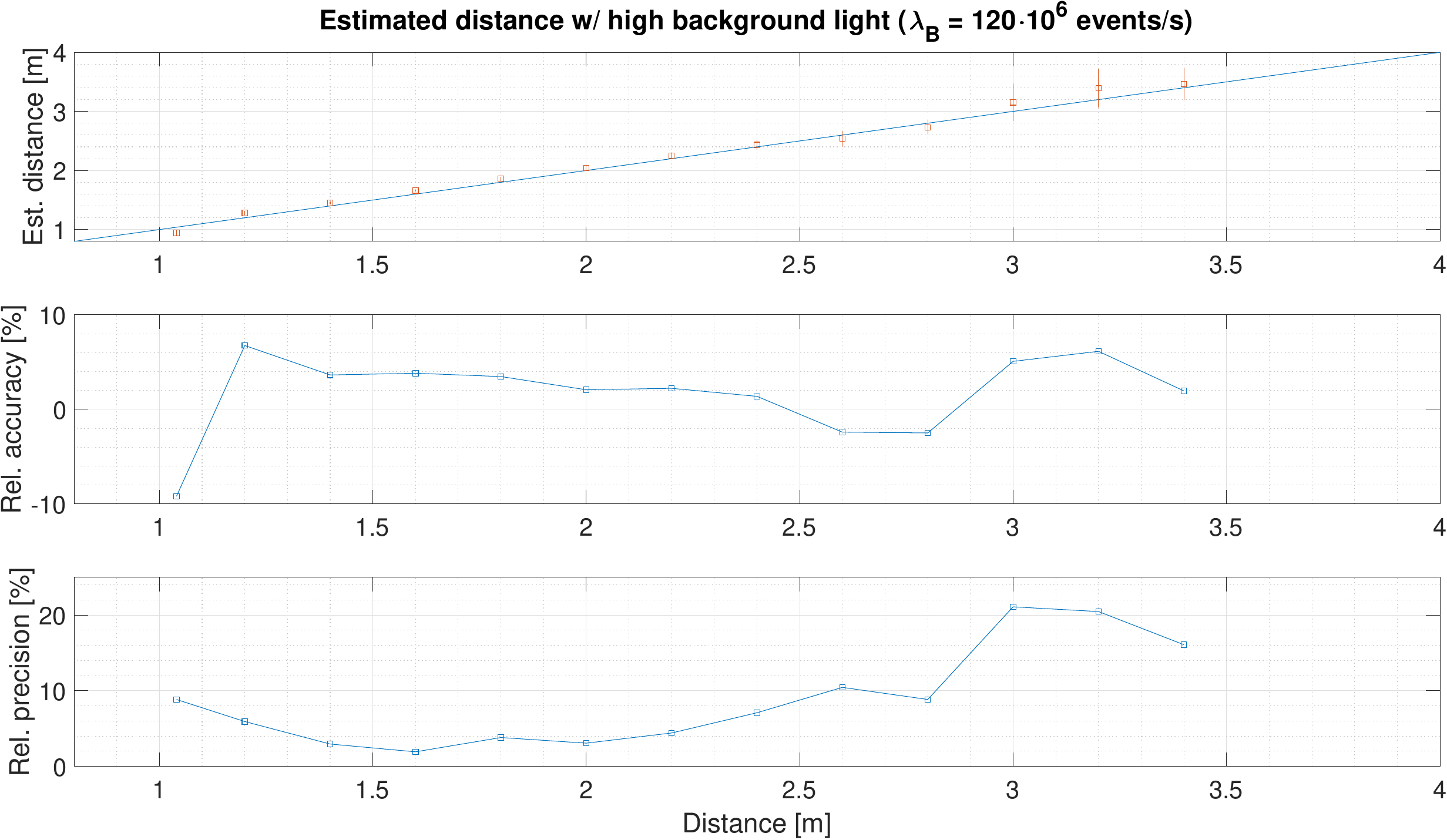}}
\caption{Measurement results with high background light flux ($\lambda_B=120\cdot10^6$~events/s), showing the
  extracted \Tof\ without the need to build a histogram of timestamps. The relative accuracy is in the range [-9,6.7]~\%, while the worst relative precision is $21\%$ at 3~m. With this high background light flux (corresponding to $\approx75$~kilolux), and the decision to use a low reflectivity target ($\approx10\%$), the maximum achieved range decreased to 3.4~m.}
\label{ToFhighBG}
\end{figure}

The last set of measurements shows the extracted \Tof\ without the need to
build a histogram of timestamps.
For each measured distance, we run the linearization algorithm 250 times to
have sufficient statistics to compute accuracy and precision.
For each run of the algorithm, we average the results from $N=1.5\cdot10^4$
vectors of linearized SPAD timestamps, to emulate an equivalent 30~FPS
operation rate, as outlined in Section~\ref{subsec:acq_description} and with
Figure~\ref{speedComparison}.
For all measurements, the same $\simeq10\%$ reflectivity target (black matte panel) was used, in the range from 1~m to 3.8~m, to emulate a challenging scenario for a typical SPAD-based d-ToF system. 
First, we evaluate the behavior of the \Tof\ extraction process without
background light.
The results, depicted in Figure~\ref{ToFnoBG}, show good agreement between the
extracted \Tof\ and the ground truth.
Then, we repeat the measurements with the inclusion of background light by
setting the halogen illuminator to generate a background light flux of
$7.7\cdot10^6$~events/s and $120\cdot10^6$~events/s.
The values of background light flux are considered at the output of the SPADs
of the sensor, and they correspond to an illumination level of
$\simeq$~15~kilolux and $\simeq$~75~kilolux, respectively.
Results are shown in Figure~\ref{ToFlowBG} and~\ref{ToFhighBG}, demonstrating
the validity of the proposed histogram-less \Tof\ estimation in a real setup.

\section{Conclusion}
\label{sec_conclusion}

In this work, we demonstrate how to extract the time-of-flight information in a SPAD-based direct time-of-flight system without the need to build a resource and bandwidth-hungry histogram of timestamps. Moreover, the proposed method is resistant against high photon fluxes and can withstand detection rates three orders of magnitude higher than the conventionally recognized limit of 5\%. The acquisition method, which is based on the linearization of the SPAD response, is
suitable for integration in CMOS technology using low resources and is
therefore scalable to large arrays, since it can be easily integrated per-pixel.
The proposed extraction method has been completely characterized, first with
Monte Carlo numerical simulations.
The method is also mathematically justified, and we demonstrated its validity
with real measurements, by repurposing an existing d-ToF sensor and using real
data to extract the ToF.
The proposed extraction method can be implemented at least in two ways, by
means of the \emph{acquire or discard} or \emph{time-gated} detection schemes.
While the \emph{acquire or discard} scheme allows for the least usage of
resources, it suffers from long integration times especially when the flux of
photons is too high.
On the other hand, the \emph{time-gated} scheme can guarantee a more efficient
acquisition at the expense of a per-pixel controllable delay element.
Concerning the ToF extraction method, we demonstrated its validity by using an
extremely low amount of resources, as only two counters and one accumulator
are required.

\clearpage

\section*{Conflicts of interest}
The authors declare no conflicts of interest.

\bibliography{sample.bib}

\end{document}